\documentclass[a4paper,11pt]{article}
\pdfoutput=1 

\usepackage{jheppub} 

\usepackage[T1]{fontenc} 
\usepackage{xcolor,colortbl}
\usepackage{subfigure}
\usepackage{graphicx}
\usepackage[compat=1.1.0]{tikz-feynman}
\usepackage{physics}
\usepackage{slashed}
\title{\boldmath 
Boosting the production of sterile neutrino dark matter with self-interactions}

\author{Mar\'{\i}a Dias Astros}
\author{and Stefan Vogl}

\affiliation{Institute of Physics, University of Freiburg,\\Herrmann-Herder-Str.~3, 79104 Freiburg, Germany}

\emailAdd{maria.dias@physik.uni-freiburg.de}
\emailAdd{stefan.vogl@physik.uni-freiburg.de}

\abstract{Sterile neutrinos are well-motivated and simple dark matter (DM) candidates. However, sterile neutrino DM produced through oscillations by the Dodelson-Widrow mechanism is excluded by current $X$-ray observations and bounds from structure formation. One minimal extension, that preserves the attractive features of this scenario, is self-interactions among sterile neutrinos. In this work, we analyze how sterile neutrino self-interactions mediated by a scalar affect the production of keV sterile neutrinos for a wide range of mediator masses. We find four distinct regimes of production characterized by different phenomena, including partial thermalization for low and intermediate masses and resonant production for heavier mediators. We show that significant new regions of parameter space become available which provide a target for future observations.}

\begin{document} 
\maketitle
\flushbottom
\section{Introduction}
\label{sec: introduction}

Sterile neutrinos are arguably one of the most minimal extensions of the Standard Model (SM). 
Despite their simplicity, the addition of fermionic singlets to the field content of the SM allows for a surprisingly large range of phenomena and they could play a role in solving several of the big puzzles of modern physics. 
For example, they can explain the origin and the smallness of neutrino masses \cite{Mohapatra:1979ia,Minkowski:1977sc,Yanagida:1979as,Gell-Mann:1979vob}, the generation of the observed baryon asymmetry via leptogenesis \cite{Luty:1992un,Akhmedov:1998qx,Davidson:2008bu,Shaposhnikov:2006nn} and the dark matter (DM) in the universe \cite{Dodelson:1993je,Shi:1998km}.  
Note, however, that these 
solutions point towards radically different mass scales and coupling strengths such that more sterile neutrinos need to be realized in nature if all of them are explained in this fashion. Here, we focus on one aspect only: sterile neutrinos as dark matter.
In the simplest case, non-thermal sterile neutrino production in the early universe is sourced by their mixing with the active neutrinos. This is known as the Dodelson-Widrow (DW) mechanism~\cite{Dodelson:1993je}. 
In this scenario, each active neutrino can be converted into a sterile state with a small probability by oscillations. In this way, one obtains the correct DM relic abundance for appropriate mixing angles.  
This setting is highly predictive since only two parameters, the sterile neutrino mass $m_s$ and the mixing angle $\theta$, 
completely control the dynamics.
The observed dark matter abundance of $\Omega h^2 = 0.12$ \cite{Planck:2018vyg} points towards $m_s= \mathcal{O}$(keV) and quite small mixing angles.  
However, this predictivity also makes the DW mechanism rather inflexible, and nowadays it is in very severe tension with observational constraints. Concretely, the mixing allows the steriles to decay into an active neutrino and a photon. Even though the lifetime is significantly longer than the age of the universe, this decay leads to a detectable signal at current and future $X$-ray telescopes \cite{Boyarsky:2007ge,Horiuchi:2013noa,Roach:2019ctw,Foster:2021ngm, Malyshev:2020hcc,Dekker:2021bos,Ando:2021fhj}. The non-detection of such a signal constrains the mixing angle and, combined with limits derived from structure formation \cite{Garzilli:2019qki,Dekker:2021scf,Hsueh:2019ynk,Gilman:2019nap,Irsic:2017ixq,Bode:2000gq}, excludes all of the parameter space where DW production allows to explain the entirety of the observed DM abundance. 

Given the alluring simplicity of sterile neutrino DM, alternative ideas for their production are highly sought after. In addition to the well-known resonant production by the Shi-Fuller mechanism \cite{Shi:1998km}, which relies on a large lepton asymmetry in the universe, novel ideas have been put forward in recent years. These include, for example, active neutrino self-interactions \cite{DeGouvea:2019wpf, Kelly:2020pcy, Benso:2021hhh}, decays of SM singlet scalars \cite{Petraki:2007gq,Merle:2013wta,Adulpravitchai:2014xna,Merle:2015oja}, and dark entropy production \cite{Hansen:2017rxr} among others. Here we focus on an alternative possibility: sterile neutrino self-interactions. Such interactions have previously been studied in opposing limits and using a different methodology by \cite{Johns:2019cwc,Bringmann:2022aim}. On the one hand, \cite{Johns:2019cwc} is limited to the heavy mediator limit. Employing the full Boltzmann ansatz for the production, this reference came to rather bleak conclusions for self-interaction enhanced sterile production. On the other hand, reference 
\cite{Bringmann:2022aim}, building on an idea proposed in \cite{Bringmann:2021tjr}, analyzed the limit of light 
mediators in a simplified framework with promising results.
In this work, we go beyond these previous studies and explore a wide range of mediator masses ranging from a few keV to a few GeV with the full Boltzmann equation for oscillation-driven sterile production. We show that interactions in the dark sector lead to four different production regimes, each characterized by different and interesting phenomena. 
This opens significant new regions of parameter space for sterile neutrino dark matter that can be probed by future experiments. 

The structure of this article is as follows. In Sec.~\ref{sec: SN production} we outline the model and explain all the ingredients needed to study the production of sterile neutrinos in the early universe. In Sec.~\ref{sec: production regimes} we discuss the different production regimes and scrutinize the dynamics that characterize them.  Next, in Sec.~\ref{sec: parameter space} we turn towards the phenomenology of the model and confront the newly found regions of parameter space with observational constraints from $X$-ray satellites and structure formation. Finally, in Sec.~\ref{sec: conclusions} we present our conclusions.

\section{Sterile neutrino production}
\label{sec: SN production}

In this work we assume a `heavy' sterile Majorana neutrino $\nu_s$ that mixes with the SM neutrinos. For simplicity, we only consider mixing with the electron neutrino $\nu_e$; generalization to a case with more complex mixing is straightforward and has a minor impact on the phenomenology considered here. The mixing angle between the sterile neutrinos and $\nu_e$ is $\theta$ and, given that sterile neutrinos are much heavier than their active counterpart ($m_s \gg m_a$), the corresponding vacuum oscillation frequency is $\omega(p) \approx \frac{m_s^2}{2 p}$. In addition, we assume that sterile neutrinos couple to a real scalar $\phi$, also a singlet under the SM gauge group, via a Yukawa interaction 
\begin{equation}
    \mathcal{L}_{\text{int}} = y~ \bar{\nu}_s \nu_s \phi.
\end{equation}
Note that we take the scalar mass $m_{\phi} \gtrsim 10~ m_s$ throughout. This allows us to neglect the mass of the sterile neutrinos and treat them as relativistic particles during production. Moreover, 
we assume that there is no sterile neutrino population in the universe prior to their production through the mechanism explained in the remainder of this section. 

The evolution of the sterile neutrinos' phase space density follows a Boltzmann equation\footnote{In full generality, the evolution of a mixed state such as the one considered here should be described by the quantum Liouville equation which studies the evolution of a density matrix. In the context of neutrino mixing in the early Universe where only certain entries of the density matrix are of interest this
goes under the label of quantum kinetic equations (QKE), see e.g. \cite{McKellar:1992ja, Sigl:1993ctk}. The Boltzmann equation Eq.~(2.2) is a classical approximation for the full QKE were some of the corrections are incorporated into an effective sterile-active transition probability. This works  well for small mixing angles, see \cite{Kishimoto:2008ic, Bell:1998ds, Johns:2019hjl} for a detailed discussion.} \cite{Foot:1996qc,DiBari:1999ha,Johns:2019cwc}
\begin{equation}
     \frac{\partial f_s(t,p)}{\partial t}-Hp\frac{\partial f_s(t,p)}{\partial p} = \frac{\Gamma_t}{4} \left< P_m(\nu_{a} \leftrightarrow \nu_s)\right>\left[f_a(t,p) - f_s(t,p) \right] + \mathcal{C}_s,
     \label{eq: Boltzmann equation}
\end{equation}
where $f_s$ and $f_a$ are the distribution functions of sterile and active neutrinos  respectively and $H$ is the Hubble rate. 
Pauli blocking has been neglected here since we are interested in the situation where the sterile neutrino gas is dilute. 
Note that all the quantities above depend on the magnitude of the three-momentum of the neutrinos, i.e. there is a Boltzmann equation for each mode. 
$\Gamma_t = \Gamma_a + \Gamma_s$ is the total interaction rate and\begin{equation}
   \left< P_m(\nu_{a} \leftrightarrow \nu_s)\right> = \frac{\omega^2(p) \sin^2(2 \theta)}{\omega^2(p) \sin^2(2\theta) + D^2(p) + \left[\omega(p) \cos(2\theta) - V_{\text{eff}}\right]^2},
   \label{eq: transition probability}
\end{equation} is the active-sterile transition probability. This probability depends on the mixing angle between the active and the sterile neutrinos and the interactions of the neutrinos with the medium. The in-medium effects are encapsulated in the quantum damping term $D(p) = \frac{\Gamma_t (p)}{2}$ and the effective potential $V_{\text{eff}}$. 
The latter is induced by forward scattering of neutrinos in the primordial plasma and has two main contributions: one piece corresponding to the SM interactions and one piece from the sterile neutrinos interactions that are mediated by $\phi$. The potential is thus given by \cite{Dasgupta:2013zpn}
\begin{equation}
    V_{\text{eff}}= V_a - V_s,
\end{equation}
where $V_{a}$ ($V_s$) is the potential seen by the active (sterile) neutrinos. The active potential reads \cite{Notzold:1987ik}
\begin{equation}
    V_a(p) = -\frac{8 \sqrt{2} G_F p }{3m_Z^2}\left( \rho_{\nu_e} + \rho_{\Bar{\nu}_e}\right) - \frac{8\sqrt{2} G_F p}{3 m_W^2} \left(\rho_e + \rho_{\Bar{e}}\right),
\end{equation}
with $G_F$ the Fermi constant, $\rho_i$ the energy density of particle species $i$ and $m_Z$ and $m_W$ the masses of the $Z$ and the $W$ bosons, respectively. Here, and in the following, we assume that the lepton asymmetry is comparable to the baryon asymmetry such that potentials proportional to $n_{\nu} - n_{\bar{\nu}}$ are negligible.
The potential for the sterile neutrinos depends on the mass of the mediator. Under certain assumptions that are discussed in Appx.~\ref{app: Effective Potential}, the expression for $V_s$ is
\begin{align}
    V_s(p) &=  - \frac{y^2}{2 p^2}
 \int_0^\infty \frac{\dd{k}}{8 \pi^2} \left[\left( \frac{m_{\phi}^2}{2}  \left[\log\left(\frac{m_{\phi}^2   + 4 k p}{m_{\phi}^2 } \right) + \log \left(\frac{-m_{\phi}^2}  {-m_{\phi}^2   + 4 k p}\right) \right]  -  4 k p  \right) f_s(k) \right. \nonumber \\ 
 & + \left. \left(  \frac{  m_{\phi}^2}{2}  \frac{k}{\epsilon_2}\left[ \log\left(\frac{m_{\phi}^2 + 2 \epsilon_2 p + 2pk}{m_{\phi}^2 + 2\epsilon_2 p - 2pk}\right) + \log\left(\frac{m_{\phi}^2 - 2 \epsilon_2 p +2pk}{m_{\phi}^2 - 2\epsilon_2 p - 2pk}\right)\right]  - \frac{4 p k^2}{\epsilon_2} \right) f_{\phi}({\epsilon_2}) \right],
 \label{eq: veff }
\end{align}
with $\epsilon_2 = \sqrt{k^2 +m_{\phi}^2}$ and the respective phase space density $f_{\phi}$ for the $\phi$ boson. In general, the integration appearing in the potential in \eqref{eq: veff } must be performed numerically. One key aspect we want to mention already here is that the potential changes sign when going from high ($T \gg m_{\phi}$) to low temperatures ($T\ll m_{\phi}$). This will become important later on when we discuss resonances in the production of sterile neutrinos. 

The interaction rate of the active neutrinos can be parametrized as \cite{Asaka:2006nq}
\begin{align}
    \Gamma_a (p) &= I(p,T) G_F^2 p T^4 ,
    \end{align}
where $I(p,T)$ is a coefficient that depends on momentum and the plasma temperature $T$. We use tabulated values for $I(p,T)$ from \cite{Laine_active_rate}. The sterile neutrino rate for 2 to 2 processes is in general given by
\begin{align}
    \Gamma_s(p) &= \sum _{\text{final}} \frac{1}{2 E} \int \dd{\Pi}_2 \dd{\Pi}_3 \dd{\Pi}_4 (2 \pi)^4 \delta^4(p+p_2-p_3-p_4) \abs{\mathcal{M}}^2_{\nu_s  \nu_s  \leftrightarrow \text{final}} \, f_s(p_2) , \label{eq: definition active-sterile rates}
\end{align}
where the $\dd{\Pi}_i = \frac{\dd[3]p_i}{(2\pi)^3 2 E_i}$ are Lorentz-invariant phase-space volume elements and $\abs{\mathcal{M}}^2_{\nu_s  \nu_s  \leftrightarrow \text{final}}$ is the matrix element for a pair of steriles going to a final state. In practice, only $\nu_s  \nu_s \rightarrow \nu_s  \nu_s$ scattering plays a role for the computation of $\Gamma_t$. The term $\mathcal{C}_s$ contains the collision integrals for the sterile-sterile scattering processes.
For our purposes two effects are relevant here: a) momentum exchange between the different modes and b) number changing processes such as $\nu_s \nu_s \rightarrow 2 \phi \rightarrow 4 \nu_s$.
The second one is higher order in $y$ and will only be relevant in certain regions of the parameter space. Therefore, we postpone its discussion for now and revisit it in Sec.~\ref{sec: light mediator case}. 
The first one is controlled by neutrino self-scattering and the associated rate is typically large. We will exploit this in the rest of the section to considerably  reduce the complexity of our problem. 
The Feynman diagrams for the respective processes are shown in Fig.~\ref{fig: feynman diagrams for nu+nu -> nu+nu and nu+nu-> phi+phi} and
expressions for the squared matrix elements and details for the integrals required for the computation of the rates are given in Appx.~\ref{app: Sterile Rates}. 

Several terms on the right hand side of \eqref{eq: Boltzmann equation} include integrals over $f_s$ which makes this an integro-differential equation. Solving this numerically is a rather involved task. However, in the problem at hand several approximations are possible which substantially reduce the complexity. We are interested in a situation where the number density of sterile neutrinos is low. In addition, the steriles undergo frequent interactions that allow for effective momentum exchange between the different modes\footnote{For the moment, this is only an assumption. However, it can easily be checked afterwards that this is indeed the case in the regions of parameter space considered here.}. Therefore, we are dealing with a dilute interacting gas and can describe the sterile neutrinos' phase space density as a Maxwell-Boltzmann distribution 
\begin{equation}
    f_s(p) = e^{-\left(\frac{p-\mu_s}{T_s}\right)},
\end{equation}
with chemical potential $\mu_s$ and temperature $T_s$\footnote{This is not fully justified in the earliest phase of production via the DW mechanism. However, in this regime the new contributions do not yet play a role and thus the error induced by this ansatz is small.}.
\begin{figure}[b!]
\centering
\begin{tikzpicture}
\begin{feynman}
\vertex (a1) {\(\nu_{s}\)};
\vertex[below = 2cm of a1] (a2){\(\nu_s\)};
\vertex at ($(a1)!0.5!(a2) + (1, 0)$)[dot] (a3);
\vertex[right=1.5cm of a3] (a4);
\vertex[right = 3.5cm of a1] (a5){\(\nu_s\)};
\vertex[below = 2cm of a5] (a6){\(\nu_s\)};
\diagram* {
{[edges=fermion]
(a1) -- [] (a3) -- [] (a2),
},
(a3) -- [scalar, edge label=\(\phi\)] (a4),
(a4) -- [fermion] (a5), 
(a6)-- [fermion] (a4)
};
\end{feynman}
\end{tikzpicture}
\hspace*{0mm}
\begin{tikzpicture}
\begin{feynman}
\vertex (a1) {\(\nu_{s}\)};
\vertex[right= 1.5cm of a1](a3);
\vertex[right = 1.2cm of a3] (a2){\(\nu_s\)};
\vertex[below=2cm of a3] (a4);
\vertex[right = 1.2cm of a4] (a5){\(\nu_s\)};
\vertex[below = 2cm of a1] (a6){\(\nu_s\)};
\diagram* {
(a1) -- [fermion] (a3),
(a3) -- [fermion] (a2),
(a3) -- [scalar, edge label=\(\phi\)] (a4),
(a4) -- [fermion] (a5), 
(a6) -- [fermion] (a4)
};
\end{feynman}
\end{tikzpicture}
\hspace*{0mm}
\begin{tikzpicture}
\begin{feynman}
\vertex (a1) {\(\nu_{s}\)};
\vertex[right= 1.5cm of a1](a3);
\vertex[right = 1.2cm of a3] (a2){\(\phi\)};
\vertex[below=2cm of a3] (a4);
\vertex[right = 1.2cm of a4] (a5){\(\phi\)};
\vertex[below = 2cm of a1] (a6){\(\nu_s\)};
\diagram* {
(a1) -- [fermion] (a3),
(a3) -- [scalar] (a2),
(a3) -- [fermion, edge label=\(\nu_s\)] (a4),
(a4) -- [scalar] (a5), 
(a4)-- [fermion] (a6)
};
\end{feynman}
\end{tikzpicture}
\caption{Representative diagrams for the relevant sterile neutrino's processes.}
\label{fig: feynman diagrams for nu+nu -> nu+nu and nu+nu-> phi+phi}
\end{figure}
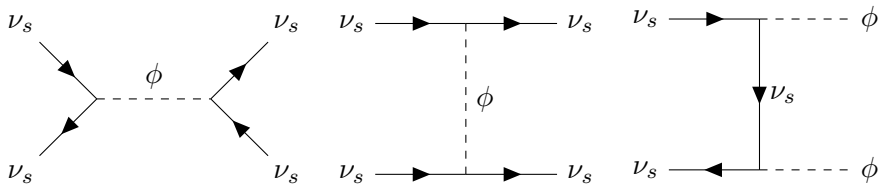Moreover, if we also assume that the process $\nu_s \nu_s \leftrightarrow \phi$ is fast enough to establish chemical equilibrium within the dark sector, \emph{i.e.} $2 \mu_s = \mu_{\phi}$, the phase space density of the scalars will be suppressed by two factors of the chemical potential
\begin{equation}
    f_{\phi}(E_\phi) = e^{-\left(\frac{E_{\phi}-\mu_{\phi}}{T_s}\right) } =e^{\left(\frac{2\mu_s}{T_s}\right) }e^{-\left(\frac{E_{\phi}}{T_s}\right) },
    \label{eq: fphi}
\end{equation}
and one can drop the terms proportional to $f_{\phi}$ in the potential in \eqref{eq: veff }.
Using the simplifications discussed above, the steriles' potential can be computed analytically in the high and low temperatures limits
\begin{equation}
   V_s(p) = \begin{cases}
        \frac{y^2 T_s^2}{2 \pi^2 p } e^{\frac{\mu_s}{T_s}} & p T_s \gg m^2_{\phi} \\[10pt]
        -\frac{16 y^2 p  T_s^4}{\pi^2 m_{\phi}^4} e^{\frac{\mu_s}{T_s}} & p T_s \ll m^2_{\phi},
        \label{eq: potential limiting cases}
    \end{cases}
\end{equation}
where the change in sign is now evident. In the intermediate regime Eq.~\eqref{eq: veff } needs to be solved numerically. Likewise, one finds an analytical expression  for the $\nu_s  \nu_s \leftrightarrow \nu_s \nu_s$ rate in these limiting cases
\begin{equation}
   \Gamma_{\nu_s  \nu_s \leftrightarrow \nu_s \nu_s}(p) = \begin{cases}
        \frac{3 y^4 T_s^2}{2 \pi^3 p } e^{\frac{\mu_s}{T_s}} & pT_s \gg m^2_{\phi} \\[10pt]
        \frac{20 y^4 p  T_s^4}{3 \pi^3 m_{\phi}^4} e^{\frac{\mu_s}{T_s}} & pT_s \ll m^2_{\phi}.
        \label{eq: rate limiting cases}
    \end{cases}
\end{equation}
For $p \, T_s =\mathcal{O}(m_\phi^2)$ the rate is dominated by the s-channel resonance. Using the narrow width approximation  leads to a simple and accurate analytic expression for the rate
\begin{equation}
\Gamma_{\nu_s \nu_s \leftrightarrow \nu_s \nu_s}^{\text{resonant}}(p) = \frac{y^2 T_s m_{\phi}^2}{2 \pi p^2}e^{-\frac{m_{\phi}^2}{4 p T_s}+\frac{\mu}{T_s}},
\end{equation}
in this regime.

Since the full phase space distribution can now be expressed by two parameters, it is convenient to turn Eq.~\eqref{eq: Boltzmann equation} into two integrated Boltzmann equations for the number density and energy density of the sterile neutrinos. They read
\begin{align}
    \dot{n}_s + 3H n_s &= \mathrm{C}_{n_s} \nonumber \\
    \dot{\rho}_s + 4H \rho_s & = \mathrm{C}_{\rho_s},
\end{align}
where the collision terms $\mathrm{C}_{n_s}$ and $\mathrm{C}_{\rho_s}$ correspond to the respective moments of the right-hand side of Eq.~\eqref{eq: Boltzmann equation}.  These are used in our numerical computations.
Because it is more convenient to work with a comoving quantity, we present our results in terms of the yield of the sterile neutrinos defined as $Y_s = n_s/s$, with $s$ the entropy density of the universe. We also track the evolution of the quantities in terms of the photon temperature~$T$.  In doing so, we make use of the SM equation of state and related thermodynamic functions of \cite{Laine:2015kra,Laine_eos}.

\section{Production regimes}
\label{sec: production regimes}
 In order to obtain the relic abundance of sterile neutrinos one needs to  solve the system of Boltzmann equations numerically. Nevertheless, one can get some useful insight into the evolution of the system by taking a closer look at the right-hand side of~\eqref{eq: Boltzmann equation}. In the classical DW mechanism, i.e. without  sterile neutrinos self-interactions or lepton-asymmetry induced resonances, 
 the peak production of $\nu_s$ happens at \cite{Dodelson:1993je}  
 \begin{equation}
     T_{\text{peak}} \sim 133~\text{MeV} ~ \left(\frac{m_s}{1~\text{keV}}\right)^{1/3}.
 \end{equation}
 This means that for keV-scale sterile neutrinos, DW production effectively stops for temperatures $T\lesssim 100 ~\text{MeV}$. Likewise, DW production of sterile neutrinos is only appreciable for $T \lesssim 1~ \text{GeV}$. The reason behind this is that at higher temperatures the SM potential and rate, which effectively scale as $T^5$, dominate over the vacuum oscillation frequency and suppress active-sterile oscillations. 
 Thus, if one wants to know the amount of sterile neutrinos produced in the early universe, one needs to integrate the Boltzmann equation starting from temperatures of a few GeV. 
 
 Adding self-interactions among sterile neutrinos can change the evolution in several ways. From the right-hand side of~\eqref{eq: Boltzmann equation} one can see that the production of sterile neutrinos strongly depends on the relative size of the rate, the vacuum oscillation frequency and the potential. This interplay is crucial and it will help us identify different regimes in the production of $\nu_s$. Let us now look into these effects in more detail. 
 
 \subsection{Light mediators}
 \label{sec: light mediator case}
 In this section we consider light mediator masses, \emph{i.e.} masses up to $\sim 100 ~\text{MeV}$.   To understand how the production of sterile neutrinos proceeds in this regime, let us revisit the relative size of the different contributions appearing on the right-hand side of \eqref{eq: Boltzmann equation}. And, for the moment, let us also drop the sterile collision term $\mathcal{C}_s$, whose impact we will address later on in the discussion. In general, at very high temperatures, the SM interactions (through the rate and the potential) will dominate, and the production of sterile neutrinos will follow DW. However, once the temperature decreases, and an initial population of sterile neutrinos is present in the plasma, self-interactions start playing a role and will change the subsequent evolution of the system.

Let us consider a typical momentum mode with $p\sim T$. 
When the mediator is light compared to the temperature 
both the sterile's potential and $\Gamma_s$ exhibit a scaling proportional to $T$, while the SM rate and potential exhibit the $T^5$ scaling behavior of heavy mediators. Thus, the importance of self-interactions increases whereas the SM contributions decrease as the temperature falls. For sufficiently low temperatures, the self-interaction rate starts to dominate. 
 On the other hand, the vacuum oscillation frequency scales as $1/T$ since $\langle p \rangle = \mathcal{O}(T)$. 
 Accordingly, for very light mediators $\omega > \Gamma_t$ and $\abs{V_{\text{eff}}}$ in the relevant temperature range\footnote{As we will see later, this holds for mediator masses up to $100~\text{MeV}$, at least for the regions of parameter space where sterile neutrinos are not overproduced.}. If the latter condition is satisfied, one can simplify the transition probability such that the Boltzmann equation can be written as \begin{equation}
    \frac{\partial f_s}{\partial t}-Hp\frac{\partial f_s}{\partial p} \sim \Gamma_t~ \sin^2(2\theta)~ f_a\,.
    \label{eq: Boltzmann equation in low mass limit}
\end{equation}
Two important facts can be deduced from this. First, the relic abundance of sterile neutrinos is clearly enhanced (compared to DW production) by the introduction of $\Gamma_s$. Second, the effective production rate over Hubble $\langle P_m \rangle \Gamma_t/H$ develops a double peak structure with a first phase of DW dominated production being followed by a second sterile interaction driven production phase. 
If these phases are clearly separated in $T$ the problem factorizes and the abundance generated by pure DW can be used to initialize a second phase described by Eq.~\eqref{eq: Boltzmann equation in low mass limit}.
This simplified ansatz for the sterile neutrino production was also put forward in \cite{Bringmann:2022aim}. We come to qualitatively similar conclusions in the regime of parameters where \eqref{eq: Boltzmann equation in low mass limit} is applicable but want to stress that this does not cover the whole range of masses and couplings 
since in-medium effects become important for heavier $m_\phi$ as we will discuss in the following subsections. In addition, we would like to remind the reader that our computation neglects the mass of the sterile neutrinos outside of the effective oscillation probability $\langle P_m \rangle$. 
This is a good approximation in the regime of parameter space considered here, i.e. $m_\phi \geq 10 m_s$, but is not expected to accurately capture the  
mass ratios $m_\phi/m_s = 3$ and $m_\phi/m_s = 5$ considered  in \cite{Bringmann:2022aim}. Therefore a direct numerical comparison with their results is not possible.

\begin{figure}[tbp]
\centering
\includegraphics[width=1\textwidth]{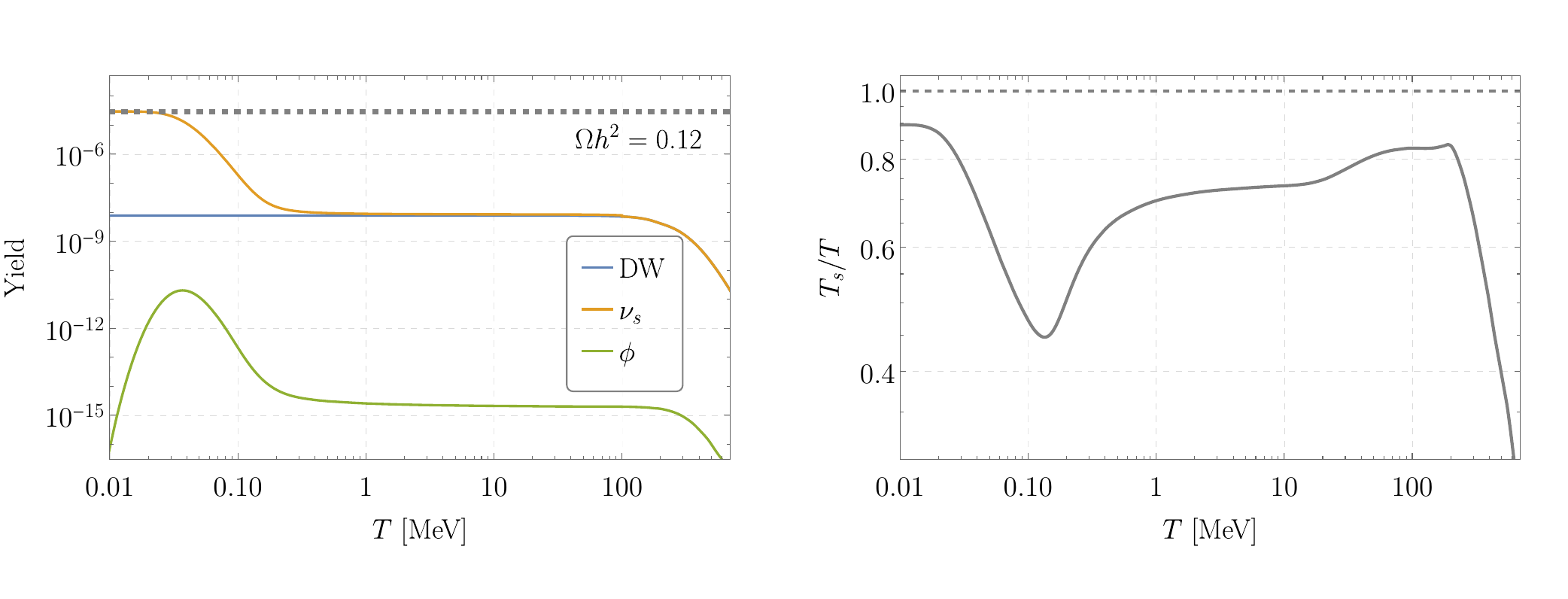}
\caption{\label{fig: light_benchmark} Evolution of the benchmark point $1$ as specified in Tab.~\ref{tab:light benchmark points}. \textbf{Left panel:} Evolution of the $\nu_s$ (orange) and $\phi$ (green) yields as a function of the photon temperature. For comparison we show the yield of sterile neutrinos produced only by DW (blue) with the same parameters, while the dashed gray line shows the observed DM relic abundance for this value of $m_s$. \textbf{Right panel:} Evolution of the temperature ratio.}
\end{figure}

To illustrate the evolution of the system in this regime, we consider a benchmark point matching the observed DM relic abundance and with a mixing angle that is small enough as to avoid current experimental constraints. The corresponding parameters for the benchmark are summarized in Tab.~\ref{tab:light benchmark points}.  
\begin{table}[tbp]
\centering
\begin{tabular}{|c|c|c|c|c|}
\hline
Benchmark&$m_s$&$m_{\phi}$&$y$&$\sin^2(2\theta)$\\
\hline 
1 & $15~\text{keV}$ & $150~\text{keV}$ & $2 \times 10^{-4}$ & $1.11 \times 10^{-13}$\\
\hline
2 & $15~\text{keV}$ & $150~\text{keV}$&$9 \times 10^{-4}$ & $2.16 \times 10^{-15}$\\
\hline
\end{tabular}
\caption{\label{tab:light benchmark points} Parameters for the two light benchmark points considered in the text.}
\end{table}
The two phases of sterile neutrino production can be seen clearly in the left panel of Fig.~\ref{fig: light_benchmark}. For  $T \gtrsim 100~\text{MeV}$ an initial population of sterile neutrinos is produced by DW.
Given that sterile neutrinos and scalars are absent at high temperatures, the sole mechanism for their production is through interactions within the SM.  
Later, the system departs from DW and, aid by self-interactions, the yield increases several orders of magnitude until it saturates at the correct value.
As can be seen in the right panel of Fig.~\ref{fig: light_benchmark}, $T_s$ remains relatively close to the temperature of the SM bath during the whole process. After DW production ends, $T_s/T$ decreases slightly since the steriles do not experience the heating from SM degrees of freedom that decouple around $T\approx100$ MeV. However, once the self-interactions become active the influx of new energy from the SM bath drives $T_s/T$ back up.

\begin{figure}[tbp]
\centering 
\includegraphics[width=\textwidth]{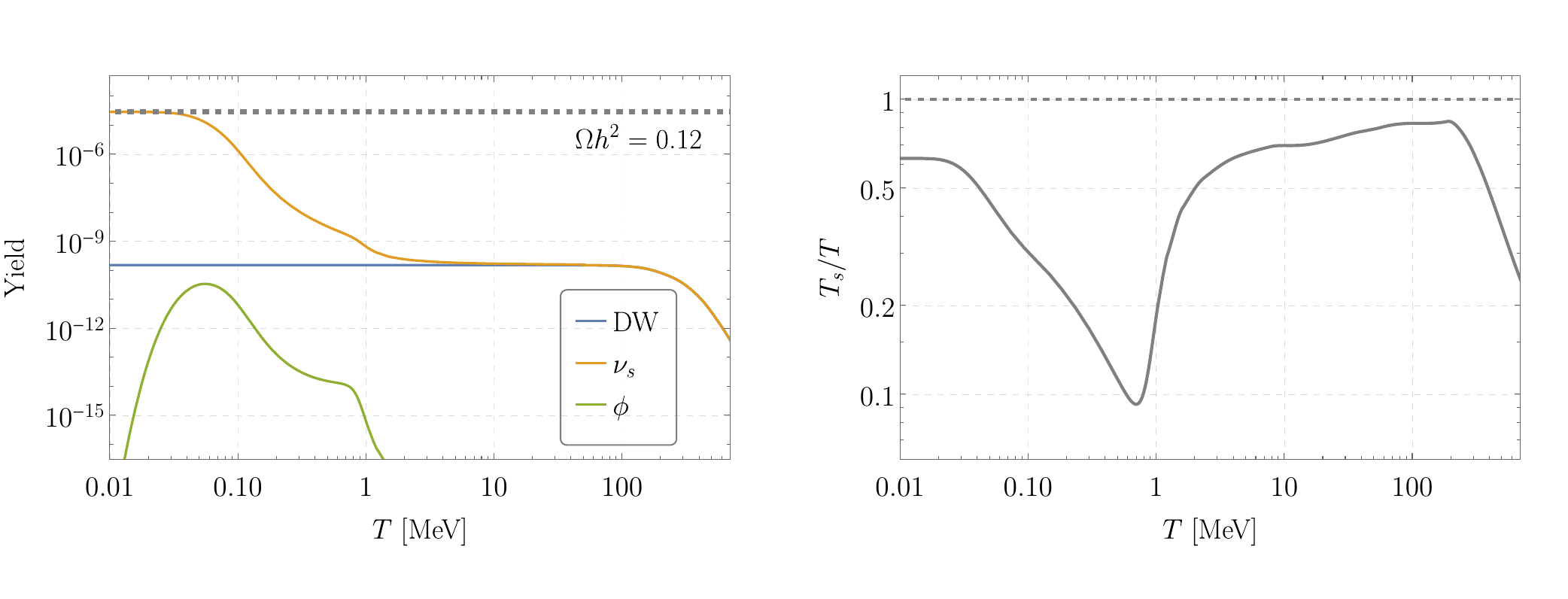}
\caption{\label{fig: thermalized benchmark}Evolution of the benchmark point $2$ as specified in Tab.~\ref{tab:light benchmark points}. \textbf{Left panel:} Evolution of the $\nu_s$ (orange) and $\phi$ (green) yields with respect to the photon temperature. For comparison we show the yield of sterile neutrinos produced only by DW (blue) with the same parameters, while the dashed gray line shows the observed DM relic abundance. \textbf{Right panel:} Evolution of the temperature ratio.}
\end{figure}

Now let us return to the discussion of the collision term $\mathcal{C}_s$. Concretely, this term accounts for the process $\nu_s \nu_s \leftrightarrow \phi \phi$ effectively resulting in $2 \nu_s \rightarrow 4 \nu_s$ upon the decay of the scalar. Assuming chemical equilibrium between $\phi$ and $\nu_s$, and  neglecting the time delay due to the finite lifetime of $\phi$, it takes the form 
\begin{equation}
    \mathcal{C}_s (p) = -\Gamma_{\nu_s \nu_s \leftrightarrow \phi \phi} \, f_s(p) \left( \frac{f_s^2 (p)}{f_{\text{eq}}^2(p)}-1 \right),
    \label{eq: number changing collision term}
\end{equation}
where $f_{\text{eq}}(p) = e^{-p/T_s}$ is the equilibrium distribution function and $\Gamma_{\nu_s \nu_s \leftrightarrow \phi \phi}$ denotes the rate for $\phi$ pair production, see Appx. \ref{app: Sterile Rates} for a more detailed discussion.
There are a few key aspects about this term. 
First, notice that $\mathcal{C}_s$ tries to drive the system towards equilibrium. 
As there is no injection of energy into the system, the increase of the sterile neutrino number density caused by the number changing interactions is associated with a cooling of the sterile neutrinos. Second, this temperature drop influences the rate and, once $T_s$ is too small, it becomes inefficient such that the thermalization process stops. Therefore, we typically only find partial thermalization and $\mu_s/T_s$ remains different from zero unless $y$ is particularly large.
The effects of this term are, in general, only important for light and intermediate mediator masses. 
To show the effects of thermalization on the system we choose a second benchmark with a higher value of $y$ and an even lower mixing angle. As depicted in the left panel of Fig.~\ref{fig: thermalized benchmark}, including the number changing process causes a deviation from DW at earlier times. 
In this case, at temperatures slightly below $\sim 10~ \text{MeV}$ the system undergoes partial thermalization resulting in a noticeable increase in the yield. 
This increase in the number of sterile neutrinos is accompanied by a sharp drop in the temperature, as shown in the right panel of Fig.~\ref{fig: thermalized benchmark}, and the process is completed shortly after its onset. The increased number density  gets amplified further by a final production phase, again sourced by the $\nu_s \nu_s \leftrightarrow \nu_s \nu_s$ scattering. As in the previous case, this leads to a new influx of energy and $T_s/T$ increases again. Note, however, that it levels of at a slightly lower value compared to the first benchmark. This reduces the impact on structure formation as will be discussed in Sec.~\ref{sec: parameter space}.

 \subsection{Intermediate mass and heavy mediators}
\label{sec: heavy mediator case}

Now we turn to the case where $m_\phi \gtrsim 100~\text{MeV}$. Here the production by DW and the one sourced by sterile-sterile interactions will in general happen simultaneously. 
Thus, the phenomenology in this regime is different compared to the case with light mediators. On the one hand, for heavy mediators number changing processes are negligible, so that we will drop the collision term $\mathcal{C}_s$ for the discussion in this section. 
On the other hand, if the mass of the mediator is large enough self-interactions can lead to resonances in the transition probability.  This can be understood by inspecting the denominator of the right hand side of Eq.~\eqref{eq: Boltzmann equation}. Under certain conditions the term in brackets in the denominator of the transition probability \eqref{eq: transition probability} can become zero.
The criterion for this to happen is
\begin{equation}
    \omega \cos(2\theta) - V_{\text{eff}} =  \omega \cos(2\theta) - V_a + V_s \approx 0\,.
    \label{eq: general resonance condition}
\end{equation}
Physically this corresponds to the $\sin^2(2 \theta_m) \rightarrow 1$ limit of the 
effective in-medium mixing angle
\begin{equation}
    \sin^2(2 \theta_m) = \frac{\omega^2(p) \sin^2(2 \theta)}{\omega^2(p) \sin^2(2\theta) + \left[\omega(p) \cos(2\theta)  - V_a + V_s \right]^2}.
\end{equation}
that occurs when the effective in-medium masses of the two neutrino eigenstates become degenerate.
When and if this happens depends on the temperature, the sterile mass, and the momentum and energy density of sterile neutrinos. The active potential is always negative during production, therefore, achieving a resonance requires the sterile potential to be negative as well. With $V_s> 0$ for $m_{\phi} \ll T_s$ this is only possible for heavy mediators.  
Once $V_s$ is negative, there are in fact two possible resonances hiding in~\eqref{eq: general resonance condition} which could boost the production of sterile neutrinos. 

On the one hand, if the creation of sterile neutrinos is rapid enough, the first resonance is achieved approximately when $\abs{V_{a}} \approx \abs{V_{s}}$ while $\omega$ is negligible. This condition is more easily met by higher energy neutrinos and at higher temperatures. More concretely the resonant criterion in this case reads
\begin{equation}
    \rho_s \approx \frac{32 \pi^3 \sqrt{2} G_F m_{\phi}^4}{y^2} \left[ \frac{(\rho_{\nu_e} + \rho_{\bar{\nu}_e})}{m_Z^2} + \frac{(\rho_e + \rho_{\bar{e}})}{m_W^2} \right], 
    \label{eq: first-resonance condition}
\end{equation}
where $\rho_s$ is the energy density of the sterile neutrinos.
This result agrees with \cite{Johns:2019cwc}.
On the other hand, a second resonance can occur at lower $T$ for $\omega = \abs{V_s}$, provided that $V_a$ is negligible at these temperatures. 
This is possible since $V_s$ can get an enhancement due to the increase of $n_s$ which changes its temperature scaling compared to $V_a$. 
It is also worth mentioning that there is a low energy cutoff $\epsilon_{min}$ below which a mode cannot pass through a resonance. 
This can easily be deduced from the fact that $\omega \rightarrow \infty$ when $p \rightarrow 0.$ 
The requirement that modes with a reasonable thermal weight need to have a negative $V_s$ points towards higher mediator masses. Empirically, we find that resonances play a role for $m_\phi\gtrsim 1$ GeV.

\begin{figure}[tbp]
\centering
\includegraphics[width=\textwidth]{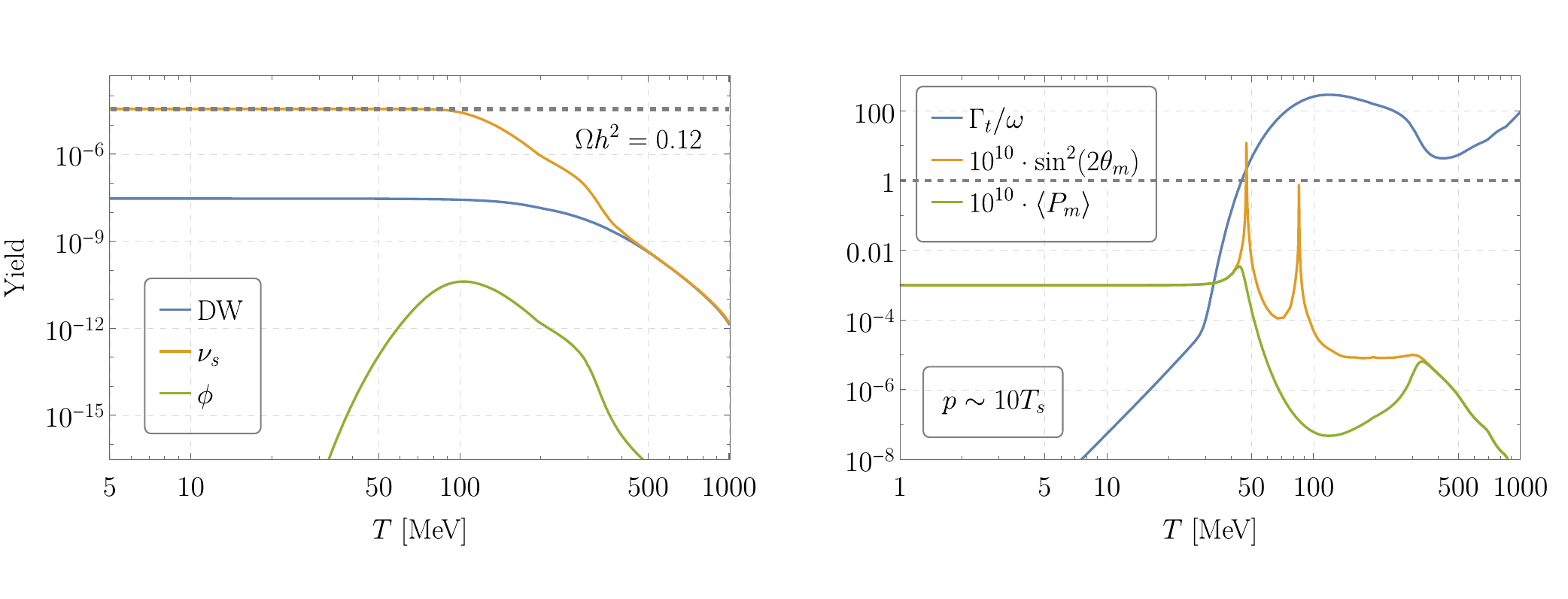}
\caption{\label{fig: heavy benchmark}Evolution of the benchmark point $3$ as specified in Tab.~\ref{tab:heavy benchmark points}. \textbf{Left panel:} Evolution of the $\nu_s$ (orange) and $\phi$ (green) yields with respect to the photon temperature. For comparison we show the yield of sterile neutrinos produced only by DW (blue) with the same parameters, while the dashed gray line shows the observed DM relic abundance. \textbf{Right panel:} Comparison of the ratio $\Gamma_t / \omega$ (blue), $\sin^2(2 \theta_m)$ (orange) and the transition probability (green) for neutrinos in the high end of the spectrum. The latter two are multiplied by a factor of $10^{10}$ to show them on the same scale.}
\end{figure}

In any case, when a resonance condition is satisfied it is easy to check that Eq.~\eqref{eq: Boltzmann equation} can be approximated as
\begin{equation}
    \frac{\partial f_s}{\partial t}-Hp\frac{\partial f_s}{\partial p} \sim \frac{\omega^2 \sin^2(2\theta)}{\Gamma_t} f_a,
    \label{eq: boltzmann equation with resonance}
\end{equation}
where we have assumed that the term $\omega^2 \sin^2(2\theta)$ in the denominator can be neglected compared to $\Gamma_t$ for the  small mixing angles considered here. 
From~\eqref{eq: boltzmann equation with resonance} one can see that the rate, which enters the denominator through the quantum damping term, plays an essential role in regulating production in this case. 
To illustrate the behavior in this regime we consider a benchmark point with a mediator mass of $1.5~\text{GeV}$ (see Tab.~\ref{tab:heavy benchmark points}), for which we select the mixing angle and the Yukawa coupling that leads to the observed DM relic abundance.
\begin{table}[tbp]
\centering
\begin{tabular}{|c|c|c|c|c|}
\hline
Benchmark&$m_s$&$m_{\phi}$&$y$&$\sin^2(2\theta)$\\
\hline 
3 & $12~\text{keV}$ & $1.5~\text{GeV}$ & $6.92 \times 10^{-2}$ & $5 \times 10^{-13}$\\
\hline
4 & $12~\text{keV}$ & $4~\text{GeV}$&$0.771 ~|~ 0.772$ & $5 \times 10^{-13}$\\
\hline
\end{tabular}
\caption{\label{tab:heavy benchmark points} Parameters for the two heavy benchmark points considered in the text.}
\end{table}
The evolution of the yield for this benchmark is depicted in the left panel of Fig.~\ref{fig: heavy benchmark} whereas
the right panel  shows a comparison between the ratio $\Gamma_t / \omega$, the transition probability and the in-medium effective mixing angle for neutrinos in the high end of the spectrum, \emph{i.e.} for which $p\sim 10 T_s$. 
As can be seen, the deviation from DW production happens at earlier times compared to lighter mediators and the sterile rate dominates $\Gamma_t$ already below $T\approx 400$ MeV. 
Around $T=50-100$ MeV  we observe that $\sin^2(2 \theta_m)$ develops a strong resonance due to the cancellation between the potentials and $\omega$. 
However, as the denominator of $\langle P_m \rangle$ is dominated by the quantum damping term, this does not affect the production and the yield remains frozen at the value it reached at $T\approx100$ MeV. Therefore, it is reasonable to call this the quantum damping regime.

\begin{figure}[tbp]
\centering
\includegraphics[width=\textwidth]{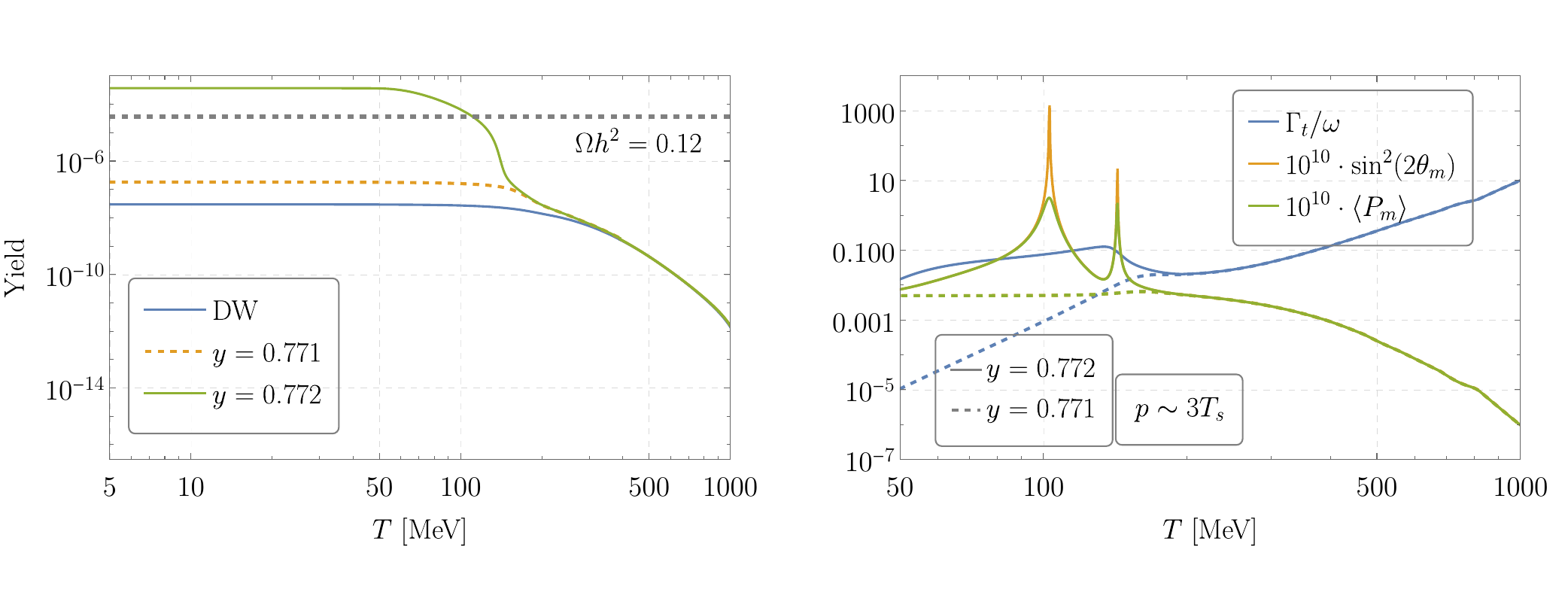}
\caption{\label{fig: fine tuned benchmark}\textbf{Left panel:} Evolution of the yield for benchmark point $4$ as specified in Tab.~\ref{tab:heavy benchmark points}. The orange curve corresponds to $y=0.771$ while the green curve is for a slightly larger $y=0.772$. \textbf{Right panel:} Comparison of the ratio $\Gamma_t/\omega$ (blue), $\sin^2(2 \theta_m)$ (orange) and the transition probability (green) for neutrinos with  a typical momentum $p\sim 3 T_s$. The latter two are multiplied by a factor of $10^{10}$ to show them on the same scale. The benchmark with lower (higher) Yukawa is shown as dashed (solid). Note that $\langle P_m \rangle$ and $\sin^2(2 \theta_m)$ overlap away from the resonances.}
\end{figure}

For even higher mediator masses, the situation is different. 
The rate, which plays an essential role in regulating the resonances, depends strongly on the ratio between temperature and mass. Especially, the resonant piece of the rate is crucial since it scales with $y^2$, unlike its heavy counterpart $\Gamma_s^{\text{heavy}}$ which scales with $y^4$. If $m_\phi^2\gg p T$ it is exponentially suppressed and the rate is dominated by the off-shell contribution $\Gamma_s^{\text{heavy}}$ which is drastically smaller. Thus the resonance in $\sin^2(\theta_m)$ is regulated less efficiently and a strong boost of the production is possible. 
Unfortunately, this is of limited use since the parameters that allow for the correct relic density turn out to be highly tuned in a large portion of the parameter space, as already pointed out in \cite{Johns:2019cwc}. 
The reason is illustrated in Fig.~\ref{fig: fine tuned benchmark} where we show the evolution of the yield for a benchmark with  $m_\phi=4~\text{GeV}$. As can be seen, a small change of the Yukawa coupling by less than one percent results in an amplification of the final yield by more than three orders of magnitude and moves the relic density from significant under- to a severe over-abundance.
This is caused by the appearance of the resonance due to the slightly larger $y$ and the slightly more efficient production at higher temperatures. Once a resonance exists, it strongly boosts the production. Therefore, a careful tuning of the Yukawa coupling against the cut-off mode $\epsilon_{min}$ is required to obtain the observed relic density here.
We do not consider such tuned solutions further.

\section{The parameter space}
\label{sec: parameter space}

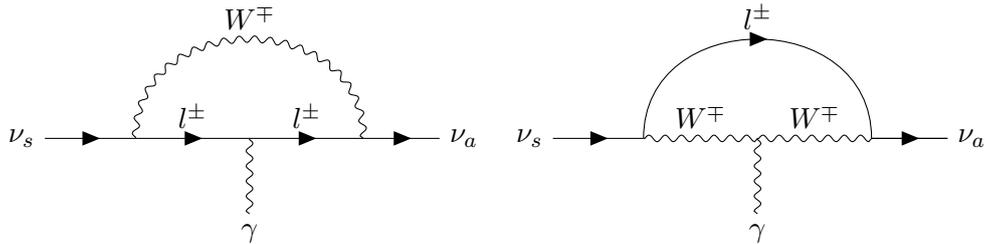
\begin{figure}[tb!]
\centering
\begin{tikzpicture}
\begin{feynman}
\vertex (a1) {\(\nu_{s}\)};
\vertex[right=1.5cm of a1] (a2);
\vertex[right=3cm of a2] (a3);
\vertex[right=1.0cm of a3] (a4) {\(\nu_{a}\)};
\vertex[right=1.5cm of a2](a5);
\vertex[below= 1cm of a5](a6) {\(\gamma\)};
\diagram* {
{[edges=fermion]
(a1) -- [] (a2) -- [edge label=\(l^{\pm}\)] (a5) -- [edge label =\(l^{\pm}\) ] (a3) -- (a4),
},
(a5) -- [boson] (a6),
(a2) -- [boson,out=90, in=90, looseness=1.5, edge label=\(W^{\mp}\), ] (a3),

};
\end{feynman}
\end{tikzpicture}
\hspace*{0mm}
\begin{tikzpicture}
\begin{feynman}
\vertex (a1) {\(\nu_{s}\)};
\vertex[right=1.5cm of a1] (a2);
\vertex[right=3cm of a2] (a3);
\vertex[right=1.0cm of a3] (a4) {\(\nu_{a}\)};
\vertex[right=1.5cm of a2](a5);
\vertex[below= 1cm of a5](a6) {\(\gamma\)};
\diagram* {
(a2) -- [boson, edge label=\(W^{\mp}\)] (a5) -- [boson, edge label =\(W^{\mp}\) ] (a3) -- (a4),
(a1) -- [fermion] (a2),
(a3) -- [fermion] (a4),
(a5) -- [boson] (a6),
(a2) -- [fermion,out=90, in=90, looseness=1.5, edge label=\(l^{\pm}\), ] (a3),

};

\end{feynman}
\end{tikzpicture}
\caption{Diagrams contributing to the radiative decay of a sterile neutrino. }
\label{fig: one-loop diagrams for SN decay}
\end{figure}

In this section, we explore the parameter space available in the model and confront it with different observational constraints. Sterile neutrinos are an example of decaying (yet cosmologically stable) DM. Due to their mixing with the SM neutrinos, the dominant decay channel for sterile neutrinos is $\nu_s \rightarrow 3 \nu_a$ which occurs at tree level. However, the one-loop process $ \nu_s \rightarrow \nu_a + \gamma$  is phenomenologically more important; see Fig.~\ref{fig: one-loop diagrams for SN decay} for a representative set of diagrams. The respective decay rate is given by \cite{Pal:1981rm}
\begin{equation}
    \Gamma_{\nu_s \rightarrow \nu_a \gamma}  \approx 1.35 \times 10^{-29} ~\text{s}^{-1} \left(\frac{\sin^2(2 \theta)}{10^{-8}}\right) \left( \frac{m_s}{1 ~ \text{keV}}\right)^5 .
\end{equation}
As this is a two-body decay the energy of the photon is $\frac{m_s}{2}$. For sterile neutrinos with $\mathcal{O}$(keV) masses, this lies in the $X$-ray band and can be searched for with current and future $X$-ray telescopes. 
These set the strongest constraints on sterile neutrino DM produced by the DW mechanism and are also very important in our scenario. 
In the following, we use a compilation of the limits presented in Refs.~\cite{Boyarsky:2007ge,Horiuchi:2013noa,Roach:2019ctw,Foster:2021ngm}.
They provide an upper limit on $\sin^2(2\theta)$ that only depends on $m_s$. 
Additionally, we need to consider structure formation bounds. 
Because sterile neutrinos are produced (and decouple) while still relativistic, they can have an impact on structure formation resembling that of warm DM (WDM), \emph{i.e} the smearing out of structures at scales that are smaller than the sterile's free-streaming length. This could then lead to a suppression in the matter-power spectrum that can be constrained with current observational data, see e.g. Refs~\cite{Dekker:2021scf,Hsueh:2019ynk, Gilman:2019nap}. In full generality, a precise evaluation of these limits requires detailed cosmological simulations. However, since the momentum distribution of our sterile neutrinos is Maxwell-Boltzmann we can match the results for thermal WDM to our case. 
The bounds on WDM are traditionally reported in terms of the smallest mass allowed for a particle with a thermal distribution and vanishing chemical potential at high temperatures. Physically, as far as the impact on structure formation is concerned, this is equivalent to a bound on the root-mean-square velocity $v_{rms}$ of the DM particle today~\cite{Bode:2000gq, Barkana:2001gr} 
\begin{equation}
    v_{rms} \approx 0.04 \left( \frac{\Omega h^2}{0.12}\right)^{1/3} \left( \frac{m_{WDM}}{1 ~\text{keV}}\right)^{-4/3} \text{km}~ \text{s}^{-1}.
\end{equation}

Hereafter, we consider a conservative limit of $m_{WDM} \gtrsim 1.9~\text{keV}$ derived in \cite{Garzilli:2019qki} from Lyman-$\alpha$ observations, which corresponds to $v_{rms} \lesssim 16~\text{m/s}$. For a Maxwell-Boltzmann distribution $v_{rms}$ can be computed easily and we get 
\begin{equation}
    v_{rms}^2 =\left.  \frac{p^2_{rms}}{m_s^2}\right|_{\text{today}}  = \left.
    \frac{1}{ n_s} \int d^3p \frac{p^2}{m_s^2} f_s(p)~   \right|_{\text{$T_d$}} \times \frac{T_0^2}{T_d^2}
    =  \frac{12}{m^2_s} \left( \frac{T_0}{T_d} T_{s,d}\right)^2,   
\end{equation}
where $T_0$ is the photon temperature today, $T_d$ is the photon temperature at the time where the sterile neutrino production is completed with corresponding temperature $T_{s,d}$. Here we used the fact that the sterile's momentum only redshifts once production is complete, \emph{i.e.} $p \propto a^{-1}$. While this is an exact statement for a decoupled species there are small corrections for a species that remains self-interacting until the non-relativistic regime. Here a small correction arises since the pressure $P$ does not transition from $P=\rho/3$ to $P\ll \rho$ instantly. Tracking the redshift evolution of $\langle p \rangle$ numerically shows that $\langle p \rangle a =$ const holds to better than $4\%$. We neglect this.

Moreover, self-interactions in the dark sector can also have an impact on structure formation, as these keep the DM particles in kinetic equilibrium after chemical decoupling~\cite{deLaix:1995vi, Atrio-Barandela:1996suw, Hannestad:2000gt}. In the presence of such interactions there is suppression of the matter-power spectrum due to pressure support, preventing the clustering of DM particles on scales smaller than the corresponding dark sound horizon. Note that the latter is related to the DM velocity dispersion in a similar way (up to order one factors) as the free-streaming length \cite{Egana-Ugrinovic:2021gnu}. The impact of self-interactions on the evolution of perturbations depends on the self-scattering cross section which  determines the time of kinetic decoupling. In the non-relativistic limit it approaches a constant value
\begin{equation}
    \sigma = 2 \frac{y^4 m_s^2}{\pi m_{\phi}^4}.
    \label{eq: self-inetraction cross section }
\end{equation}
 The impact is simple to estimate in two limiting cases. On the one hand, if kinetic decoupling happens well within the relativistic regime, DM particles will free-stream as in models of WDM, \emph{i.e.} the limit of $\sigma/m_s =0$. 
On the other hand, if self-interactions keep the DM particles in kinetic equilibrium such that they behave as a perfect fluid until the time of matter-radiation equality, DM particles do not free-stream and the structure formation bounds are therefore relaxed.  The quantitative effect of this can be estimated by realizing that the sound horizon of a strongly coupled fluid is a factor of $1/\sqrt{3}$ smaller than the free-streaming length of the same non-interacting species. This then translates to a bound on the root-mean-square velocity that is weaker by a factor of $3^{-1/4}$. In numerical studies the departure from the pure free-streaming case is found to be $\sigma /m_s \approx 10^{-5} \text{cm}^2 / \text{g}$ \cite{Egana-Ugrinovic:2021gnu} which corresponds to kinetic decoupling before the modes that are relevant for Lyman-$\alpha$ enter the horizon. For larger values the bound relaxes towards the perfect fluid limit which is reached for $\sigma /m_s \approx 1 \text{cm}^2 / \text{g}$. We do not attempt to model the bound in this intermediate regime and rather indicate the free-streaming and the perfect fluid bound as well as the onset of the transition region.

\begin{figure}[tbp]
\centering
\includegraphics[width=.95\textwidth]{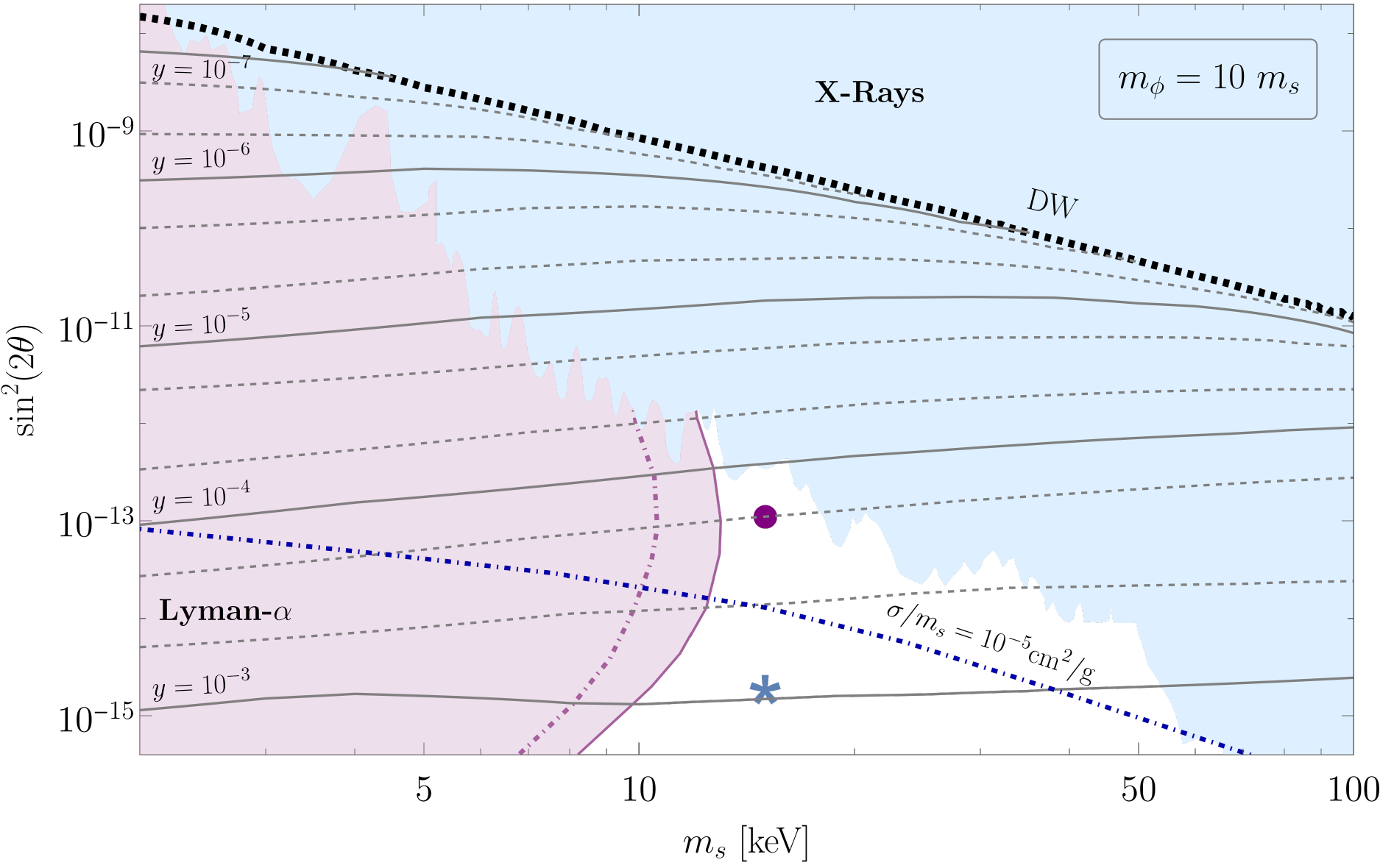}
\caption{\label{fig: parameter space} Parameter space available in the $\sin^2(2\theta)$-$m_s$ plane for a fixed mediator mass $m_{\phi}= 10~m_s$. The Yukawa contours (gray) are chosen in such a way that the full DM relic abundance is obtained everywhere in the plot (dashed gray lines correspond to intermediate values of $2$ and $5\times 10^{-x}$ ). In addition we show the DW line (dashed black) together with the compilation of $X$-rays constraints (light blue) and Lyman-$\alpha$ bounds (light purple). Points below the dash-dotted blue line satisfy $\sigma/m_s \geq 10^{-5} \text{cm}^2/\text{g}$ where interactions are expected to relax the free streaming bound. The purple dash-dotted line indicates the relaxed Lyman-$\alpha$ bounds in the perfect fluid limit, see text for more details. 
The benchmark points given in Tab.~\ref{tab:light benchmark points} are also shown for reference.}
\end{figure}

\begin{figure}[t]
\centering
\includegraphics[width=.95\textwidth]{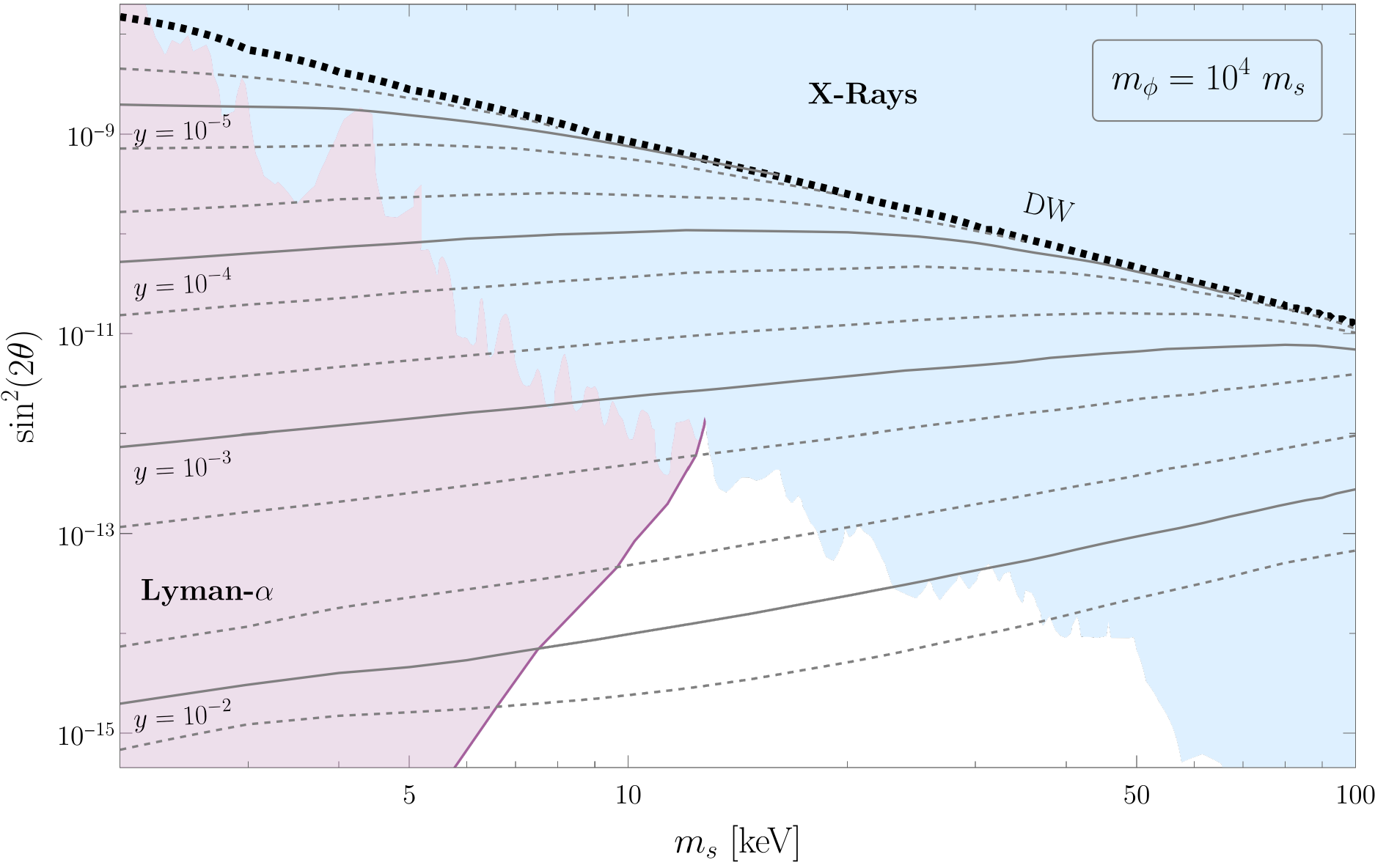}
\caption{\label{fig: heavy parameter space} Parameter space available in the $\sin^2(2\theta)$-$m_s$ plane for a fixed mediator mass $m_{\phi}= 10^4~m_s$. The Yukawa contours (gray) are chosen in such a way that the full DM relic abundance is obtained everywhere in the plot (dashed gray lines correspond to intermediate values of $2$ and $5\times 10^{-x}$ ). In addition, we show the DW line (dashed black) together with the compilation of $X$-rays constraints (light blue) and Lyman-$\alpha$ bounds (light purple). Here $\sigma/m_s \leq 10^{-5} \text{cm}^2/\text{g}$ everywhere and we do not show the relaxed Lyman-$\alpha$ bound.}  
\end{figure}

In Figs.~\ref{fig: parameter space} and \ref{fig: heavy parameter space} we show the available parameter space of the model in the $\sin^2(2\theta)$-$m_s$ plane for two exemplary ratios of masses $m_{\phi}/m_s$ = 10 and $m_{\phi}/m_s = 10^4$, respectively. For each point in the $m_s$-$\sin^2(2\theta)$-plane, the Yukawa couplings were chosen such that the full DM relic abundance is obtained. The values of $y$ are indicated by gray solid and dashed lines. As expected, one can see that for the heavier the mediator the Yukawa coupling must be larger in order to produce enough DM. 
Along the dashed black line, sterile neutrinos make up the entirety of DM via the canonical DW mechanism. 
As it is well known, pure DW is excluded, mostly due to $X$-rays. Including interactions in the sterile sector allows to fulfill the relic density requirement for mixing angles well below the DW line and opens up significant new regions of parameter space. However, the unconstrained parameter space (white regions in Figs.~\ref{fig: parameter space} and \ref{fig: heavy parameter space}) is clearly limited. 
First, constraints from $X$-ray searches bound the parameter space from above and remove the upper right corner in a way that is completely analogous to the DW scenario and which does not depend on the details of the sterile self-interactions. Second, on the left, Lyman-$\alpha$ observations set a lower limit on the mass of the sterile neutrinos. In the case of the small mass ratio, these are relatively strong since the bulk of the neutrinos is produced rather late and, therefore, they inherit the heating of the SM plasma due to the decreasing number of relativistic degrees of freedom. This contrasts with the larger mass ratio, where neutrinos are produced at earlier times and thus, the Lyman-$\alpha$ constraints become weaker compared to the case of lighter mediators. In both cases, these bounds relax somewhat towards low mixing angles since the associated larger Yukawa couplings allow for partial thermalization. This leads to a decrease of the temperature and, as a consequence, less free-streaming. In addition,  we show the region for which $\sigma /m_s \geq 10^{-5} \text{cm}^2/\text{g}$ (below dash-dotted blue line). This is visible for the smaller mass ratio and outside of the plot range for $m_\phi/m_s =10^4$. In this region one expects corrections to the Lyman-$\alpha$ bounds since we are dealing with interacting warm dark matter. This relaxes the bounds a bit. In the extreme limit, for DM that behaves as a perfect fluid until the time of matter-radiation equality, the bounds relax to the dash-dotted purple line.

\section{Conclusions}
\label{sec: conclusions} 

Sterile neutrinos are one of the most minimal extensions of the SM. Apart from having interesting implications for neutrino masses and the baryon asymmetry of the universe, they also constitute promising candidates for DM. However, the simplest realization of sterile neutrino DM is excluded by observational constraints. Therefore, it is of great interest to study possible extensions of the canonical DW scenario. In this work, we focused on sterile neutrinos with a new secret interaction mediated by a scalar with a mass in the keV to GeV range. 
The phenomena that affect DM production are surprisingly rich and we found different production regimes that are mainly defined by the mass of the mediator and the Yukawa coupling. It is hard to draw clear lines between these regimes, in particular since the evolution of the system also depends on the mixing angle. Nonetheless, as a general rule of thumb, we find that, as long as the combination of $y$ and $\theta$ is such that DM is not overproduced,  different regimes can be distinguished: 
\begin{enumerate}
  \item For $m_{\phi} \lesssim 100~\text{MeV}$ the system is well described by means of a Boltzmann equation of the form~\eqref{eq: Boltzmann equation in low mass limit}. Here, only the interaction rate matters and effects due to the potentials and quantum damping can be neglected. As the DW mechanism and the scalar interactions typically favor different temperature ranges the problem can be `factorized' into two phases.  If the Yukawa coupling is sizable one can get a partial thermalization of the system which provides a further boost to the relic density. 
 \item For mediator masses greater than $100~ \text{MeV}$ one needs to track the evolution of sterile neutrinos by fully solving~\eqref{eq: Boltzmann equation} since the production by DW and the production by self-interactions will necessarily overlap in time (or temperature). However, up to $m_{\phi} \sim 800 \text{MeV}$ the mediator is not heavy enough to allow for resonances in the effective mixing angle. Quantum damping might be important, though, at least for neutrinos at the high end of the spectrum. 
    \item For mediator masses around $1$ GeV the effective mixing angle $\sin^2(2\theta_m)$ can develop resonances for some momentum modes. However, at the same time, the on-shell part of the scattering rate is large. Thus the quantum damping term dominates and $\langle P_m \rangle \propto 1/\Gamma_t^2$ which limits production. 
    \item Finally, for $m_{\phi} \gtrsim 3~\text{GeV}$ one reaches the heavy mediator limit, where the resonances are no longer regulated effectively.  Here, the sterile neutrino abundance transitions sharply between overproduction and underproduction of DM.  
\end{enumerate}
Except for case no. 4, all of these regimes allow to produce the right amount of dark matter for a large range of masses and couplings at significantly smaller mixing angles than in the DW case. Parts of this new parameter space are already excluded by $X$-ray searches and the impact on structure formation. Nevertheless, significant new regions become viable that may be tested with more precise observations.

\acknowledgments

The work of MD and SV has been supported by the German Research Foundation (DFG) via the Individual Research Grant 496940663 and the Research Training Group (RTG) 2044.

We thank P. F. Depta for his comments and for raising a concern that made us realize an issue with our original numerical implementation in the limit where the temperature of the steriles differs substantially from the one of the actives.

\appendix

\section{Effective potential}
\label{app: Effective Potential}

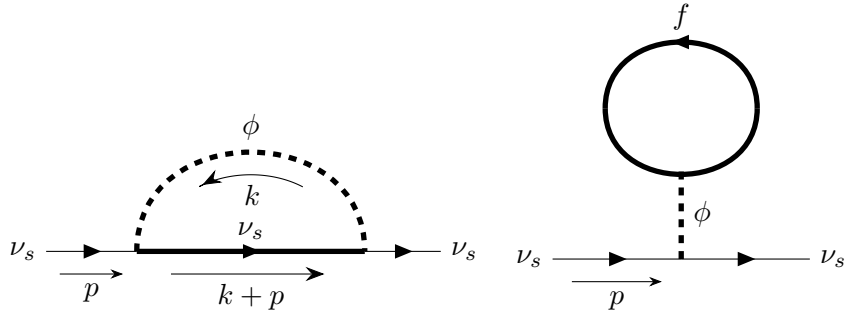
\begin{figure}[b!]
\centering
\begin{tikzpicture}
\begin{feynman}
\vertex (a1) {\(\nu_{s}\)};
\vertex[right=1.5cm of a1] (a2);
\vertex[right=3cm of a2] (a3);
\vertex[right=1.0cm of a3] (a4) {\(\nu_{s}\)};

\diagram* {
{[edges=fermion]
(a1) -- [momentum'=\(p\)] (a2) -- [edge label=\(\nu_{s}\), momentum'=\(k+p\), line width = 0.7mm] (a3) --  (a4),
},
(a2) -- [scalar, rmomentum'={[arrow shorten=0.3]\(k\)},out=90, in=90, looseness=1.5, edge label=\(\phi\), line width = 0.7mm] (a3),

};
\end{feynman}
\end{tikzpicture}
\hspace*{0mm}
\begin{tikzpicture}
\begin{feynman}
\vertex (a1) {\(\nu_{s}\)};
\vertex[right=2cm of a1] (a2);
\vertex[right=4cm of a1] (a3){\(\nu_{s}\)};
\vertex[right=1cm of a1](a4);
\vertex[right=1cm of a2](a5);

 \vertex[above=2 cm of a4](c1) ;
 \vertex[above=2cm of a5](c2);
\vertex[above=1.2cm of a2] (b1);

\diagram* {
{[edges=fermion]
(a1) -- [momentum'=\(p\)] (a2) -- (a3),
},
(a2) -- [scalar,edge label'=\(\phi\), line width = 0.7 mm] (b1),
(c2) -- [fermion, edge label'=\(f\), out=90, in=90, looseness=1.5, line width = 0.7mm] (c1),
(c1) -- [plain, out=-90, in=-90, looseness=1.5, line width = 0.7mm] (c2),

};

\end{feynman}
\end{tikzpicture}
\caption{Lowest order contributions to the sterile neutrinos' self-energy. Thick lines represent thermal propagators.}
\label{fig: one-loop diagrams for veff}
\end{figure}

As briefly discussed in the main text, the interactions of neutrinos with the plasma are responsible for a modification in their dispersion relation through an effective potential, see e.g. \cite{Notzold:1987ik,Quimbay:1995jn,Dasgupta:2013zpn,Jeong:2018yts}
\begin{equation}
    E = \abs{\bold{p}} + \frac{m^2}{2\abs{\bold{p}}} + V_{\text{eff}}. 
    \label{eq: modified dispersion relation}
\end{equation}
This is analogous to regular neutrino oscillations where matter effects can have a significant impact on the observed oscillations. Indeed, this potential encapsulates thermal corrections to the neutrinos' self-energy. The lowest order of such corrections appear at one loop level in the bubble and tadpole diagrams shown in Fig.~\ref{fig: one-loop diagrams for veff}, where the thick lines represent thermal propagators.
One can compute these diagrams in Feynman gauge in the real-time formalism and we proceed in this way in this section, closely following \cite{Quimbay:1995jn}. We also consider ultra-relativistic neutrinos for the calculation. The tadpole diagram gives a contribution that is proportional to the fermion asymmetry and, since we assume this asymmetry to be comparable to the baryon asymmetry, it can be neglected. 
For the bubble diagram the self energy correction is given by 
\begin{equation}
    \Sigma(p) = i y^2 \int \frac{\dd[4]{k}}{(2\pi) ^4} D(k) S(p+k), 
    \label{eq: one-loop self-energy correction}
\end{equation}
with the thermal propagators 
 \begin{align}
     S(k) &= \slashed{k} \left[ \frac{1}{k^2  + i \epsilon} + i \Gamma_f(k) \right] \\ 
     D(k)&=\frac{1}{k^2 - m_{\phi}^2+i\epsilon}- i\Gamma_b(k),
 \end{align}
 and the corresponding thermal functions 
 \begin{align}
     \Gamma_f(k) &= 2 \pi \delta(k^2) f_f(k \cdot u) \\ 
     \Gamma_b(k) &= 2 \pi \delta(k^2 - m_{\phi}^2) f_b(k \cdot u).
 \end{align}
In the same way, the fermionic and bosonic distribution functions are in general of the form\footnote{The distribution function can also be a Maxwell-Boltzmann distribution as specified in the main text.} 
\begin{align}
f_f (k \cdot u ) &= \left[e^{\abs{k \cdot u }/T_s} +1\right]^{-1}  \\
f_b (k \cdot u ) &= \left[e^{\abs{k \cdot u }/T_s} -1\right]^{-1}, 
\end{align}
where $u^{\mu}$ is the four-velocity of the plasma. We choose to work in the  rest frame of the heat bath so we take $u = (1,0,0,0).$ The effective potential appearing in \eqref{eq: modified dispersion relation} can then be computed as 
 \begin{equation}
     V_{\text{eff}} = -\frac{1}{4 \abs{\bold{p}}^2} \left[\left((p^0)^2-\abs{\bold{p}}^2\right) \Tr(\slashed{u} \Re{\Sigma(p)}) - p^0 \Tr(\slashed{p} \Re{\Sigma(p)})\right],
     \label{eq: general form Veff}
 \end{equation}
so we need to evaluate the expression in~\eqref{eq: one-loop self-energy correction}. In order to do that we first take the leading thermal correction, \emph{i.e.} the terms in~\eqref{eq: one-loop self-energy correction} that are proportional to one power of $\Gamma_f$ or $\Gamma_b$ 
\begin{align}
    \Re{\Sigma(p)} &= y^2  \int \frac{\dd[4]{k}}{(2\pi) ^4} \left(\slashed{p} + \slashed{k}\right) \left[ \frac{\Gamma_b (k)}{(p +k)^2} - \frac{\Gamma_f(p+k)}{k^2 - m_{\phi}^2}\right].
    \label{eq: real part of fermionic self-energy}
\end{align}
Let us now focus on the first term of the integral which in expanded form reads 
\begin{align}
    I_1 = \int \frac{\dd[4]{k}}{(2\pi) ^4} \left(\slashed{p} + \slashed{k}\right)  \frac{\Gamma_b (k)}{(p +k)^2} =  \int \frac{\dd[4]{k}}{(2\pi) ^4} \left(\slashed{p} + \slashed{k}\right) \frac{2 \pi \delta( k^2 - m_{\phi}^2) f_b(k \cdot u)}{(p+k)^2}.
    \label{eq: first term self-energy}
\end{align}
Keeping in mind that $\Tr(\slashed{a}\slashed{b}) = 4 ~ (a \cdot b)$ we can take the trace and get  
 \begin{align}
     \frac14 \Tr{\slashed{u}I_1} &= \int \frac{\dd[4]{k}}{(2\pi) ^3} \left[ u \cdot \left(p+k \right) \right]\frac{\delta( k^2 - m_{\phi}^2) f_b(k \cdot u)}{(p+k)^2} \nonumber \\ 
   &=  \int \frac{\dd[4]{k}}{(2\pi)^3}\frac{\left(p^0+k^0\right)}{(p+k)^2} \frac{f_b (k^0)}{2 \sqrt{\abs{\bold{k}}^2 + m_{\phi}^2}} \left[\delta\left(k^0 - \sqrt{\abs{\bold{k}}^2+m_{\phi}^2}\right) +\delta\left(k^0 + \sqrt{\abs{\bold{k}}^2+m_{\phi}^2}\right)\right],
 \end{align}
where we have used the fact that \begin{equation*}
    \delta( k^2 - m_{\phi}^2) = \frac{1}{2 \sqrt{\abs{\bold{k}}^2 + m_{\phi}^2}} \left[\delta\left(k^0 - \sqrt{\abs{\bold{k}}^2+m_{\phi}^2}\right) +\delta\left(k^0 + \sqrt{\abs{\bold{k}}^2+m_{\phi}^2}\right)\right].
\end{equation*}
By adopting the notation from \cite{Quimbay:1995jn} where $\epsilon_2 = \sqrt{\abs{\bold{k}}^2 + m_{\phi} ^2}$ we can further simplify our expression using the delta functions to perform the integral over $k^0$. The result reads
\begin{align}
 \frac14 \Tr{\slashed{u}I_1} &= \frac12 \int \frac{\dd[3]{\bold{k} }}{(2\pi)^3} \left[ \frac{p^0 + \epsilon_2}{(p+k)^2} +\frac{p^0 - \epsilon_2}{(p+k)^2} \right] \frac{f_b (\epsilon_2)}{\epsilon_2}\nonumber \\ &= \frac12 \int \frac{\dd{\abs{\bold{k}}} \dd{\cos(\theta)} \abs{\bold{k}}^2}{(2\pi)^2} \left[ \frac{p^0 + \epsilon_2}{p^2 + m_{\phi}^2 + 2 p^0 \epsilon_2 - 2 \abs{\bold{k}}\abs{\bold{p}} \cos(\theta)}  \right.\nonumber \\ &+ \left.\frac{p^0 - \epsilon_2}{p^2 + m_{\phi}^2 - 2 p^0 \epsilon_2 - 2 \abs{\bold{k}}\abs{\bold{p}} \cos(\theta)} \right] \frac{f_b (\epsilon_2)}{\epsilon_2},
\end{align}
where, in the second line, we have performed one of the angular integrations and denoted by $\theta$ the angle between $\bold{k}$ and $\bold{p}$. Finally, the integration over $\cos(\theta)$ is straightforward and yields 
\begin{align}
    \frac14 \Tr{\slashed{u}I_1} &= \frac12 \int \frac{\dd{\abs{\bold{k}}} \dd{\cos(\theta)} \abs{\bold{k}}^2}{(2\pi)^2} \left[ \frac{(p^0 + \epsilon_2)}{2 \abs{\bold{k}}\abs{\bold{p}} }  \log\left(\frac{p^2 + m_{\phi}^2 +2 p^0 \epsilon_2 + 2 \abs{\bold{p}} \abs{\bold{k}}}{p^2 + m_{\phi}^2 +2 p^0 \epsilon_2 - 2 \abs{\bold{p}} \abs{\bold{k}}}\right) \right.\nonumber \\ &+ \left.\frac{(p^0 - \epsilon_2)}{2 \abs{\bold{k}} \abs{\bold {p}}} \log\left(\frac{p^2 + m_{\phi}^2 -2 p^0 \epsilon_2 + 2 \abs{\bold{p}} \abs{\bold{k}}}{p^2 + m_{\phi}^2 -2 p^0 \epsilon_2 - 2 \abs{\bold{p}} \abs{\bold{k}}}\right)\right] \frac{f_b (\epsilon_2)}{\epsilon_2}.
\end{align}
The computation of the other trace appearing in~\eqref{eq: general form Veff} proceeds in a similar manner. For the evaluation of the second term in~\eqref{eq: real part of fermionic self-energy}, on the other hand, it will be convenient to use a redefinition of variables. For instance, by setting $k \rightarrow -(p+k)$, the second term reads 
\begin{align}
    I_2 = -\int \frac{\dd[4]{k}}{(2\pi) ^4} \left(\slashed{p} + \slashed{k}\right)  \frac{\Gamma_f(p+k)}{k^2 - m_{\phi}^2} = \int \frac{\dd[4]{k}}{(2\pi) ^4}  \slashed{k}  \frac{\Gamma_f(-k)}{(k+p)^2 - m_{\phi}^2}.
\end{align}
In this way, we obtain an expression that is essentially the same as the one appearing in~\eqref{eq: first term self-energy} and we can follow the steps shown above. Therefore, by putting all the pieces together into~\eqref{eq: general form Veff} one arrives at the result of \cite{Quimbay:1995jn} (by setting $m_s=0$) for the effective potential 
\begin{align}
    V_{\text{eff}} &= -\frac{y^2}{2\abs{\bold{p}}^2} \int_0^{\infty} \frac{\dd{\abs{\bold{k}}}}{8 \pi^2}\left[ \left( \frac{\left((p^0)^2 - \abs{\bold{p}}^2\right)}{\abs{\bold{p}}} \abs{\bold{k}} L_1^-(\abs{\bold{k}}) - \frac{\left((p^0)^2 - \abs{\bold{p}}^2 - m_{\phi}^2\right)}{2} \frac{p^0}{\abs{\bold{p}}}  L_1^+(\abs{\bold{k}}) - 4 p^0 \abs{\bold{k}}\right) f_f(\abs{\bold{k}}) \right. \nonumber \\ &+ \left. \left( \frac{\left((p^0)^2 - \abs{\bold{p}}^2\right)}{\abs{\bold{p}}} \abs{\bold{k}} L_2^-(\abs{\bold{k}}) + \frac{\left((p^0)^2 - \abs{\bold{p}}^2 + m_{\phi}^2\right)}{2} \frac{p^0}{\abs{\bold{p}}} \frac{\abs{\bold{k}}}{\epsilon_2} L_2^+(\abs{\bold{k}}) - \frac{4 p^0 \abs{\bold{k}}^2}{\epsilon_2} \right) f_b({\epsilon_2}) \right],
    \label{eq:integral version of Veff}
\end{align}
with the logarithmic functions defined as 
\begin{align}
    L_1^{\pm}(\abs{\bold{k}}) &= \log\left[\frac{(p^0)^2 - \abs{\bold{p}}^2 - m_{\phi}^2 - 2\abs{\bold{k}} p^0 - 2 \abs{\bold{p}} \abs{\bold{k}}}{(p^0)^2 - \abs{\bold{p}}^2 - m_{\phi}^2 - 2\abs{\bold{k}} p^0 + 2 \abs{\bold{p}} \abs{\bold{k}}}\right] \pm \log\left[\frac{(p^0)^2 - \abs{\bold{p}}^2 - m_{\phi}^2 + 2\abs{\bold{k}} p^0 - 2 \abs{\bold{p}} \abs{\bold{k}}}{(p^0)^2 - \abs{\bold{p}}^2 - m_{\phi}^2 + 2\abs{\bold{k}} p^0 + 2 \abs{\bold{p}}  \abs{\bold{k}}}\right] \nonumber \\ \nonumber \\  L_2^{\pm}(\abs{\bold{k}}) &= \log\left[\frac{(p^0)^2 - \abs{\bold{p}}^2 + m_{\phi}^2 + 2\epsilon_2 p^0 + 2 \abs{\bold{p}} \abs{\bold{k}}}{(p^0)^2 - \abs{\bold{p}}^2 + m_{\phi}^2 + 2\epsilon_2 p^0 - 2 \abs{\bold{p}} \abs{\bold{k}}}\right] \pm \log\left[\frac{(p^0)^2 - \abs{\bold{p}}^2 + m_{\phi}^2 - 2 \epsilon_2 p^0 + 2 \abs{\bold{p}} \abs{\bold{k}}}{(p^0)^2 - \abs{\bold{p}}^2 + m_{\phi}^2 - 2\epsilon_2 p^0 - 2 \abs{\bold{p}}  \abs{\bold{k}}}\right].
\end{align}
 The integral appearing in~\eqref{eq:integral version of Veff} must be solved numerically. Nonetheless, there are a couple of simplifications that we can do beforehand. The first simplification is to assume that $V_{\text{eff}} \ll \abs{\bold{p}}$ such that we can replace $p^0 \approx \abs{\bold{p}}$ directly into $V_{\text{eff}}$. With this in mind one gets 
 \begin{align}
     V_{\text{eff}} &= -\frac{y^2}{2\abs{\bold{p}}^2} \int_0^{\infty} \frac{\dd{\abs{\bold{k}}}}{8 \pi^2}\left[ \left( \frac{  m_{\phi}^2}{2}  L_1^+(\abs{\bold{k}}) - 4 \abs{\bold{p}} \abs{\bold{k}}\right) f_f(\abs{\bold{k}}) + \left(  \frac{  m_{\phi}^2}{2}  \frac{\abs{\bold{k}}}{\epsilon_2} L_2^+(\abs{\bold{k}}) - \frac{4 \abs{\bold{p}} \abs{\bold{k}}^2}{\epsilon_2} \right) f_b({\epsilon_2}) \right].
     \label{eq: simplified version of Veff}
 \end{align}
Furthermore, as it is the case with the rate, one can easily compute the heavy and the massless limits for the potential. First, let us consider the high-temperature limit, \emph{i.e.} $\abs{\bold{p}}, T_s\gg m_{\phi}$. In this case we can drop the sub-leading logarithmic terms and get 
\begin{align}
    V_{\text{eff}}(\abs{\bold{p}},T_s \gg m_{\phi}) &=   \frac{y^2}{2\abs{\bold{p}}^2} \int_0^{\infty} \frac{\dd{\abs{\bold{k}}}}{8 \pi^2}\left[ 4 \abs{\bold{p}} \abs{\bold{k}} f_f(\abs{\bold{k}}) + \frac{4 \abs{\bold{p}} \abs{\bold{k}}^2}{\epsilon_2}  f_b({\epsilon_2}) \right] \nonumber \\ &= \frac{2 y^2}{\abs{\bold{p}}} \int_0^{\infty} \frac{\dd{\abs{\bold{k}}} \abs{\bold{k}}}{8 \pi^2}\left[  f_f(\abs{\bold{k}}) +   f_b(\abs{\bold{k}}) \right], 
\end{align}
 where in the second line we have used the fact that for high temperatures, $\epsilon_2 \approx \abs{\bold{k}}$. Hence, by assuming Fermi-Dirac and Bose-Einstein distribution functions, the high $T$ limit of the potential is just 
 \begin{equation}
 \label{eq:pot1}
      V_{\text{eff}}(\abs{\bold{p}},T_s \gg m_{\phi}) = \frac{y^2}{8 \abs{\bold{p}}} T_s^2. 
 \end{equation}
Conversely, for the low temperature limit, one needs to expand the potential at leading order in $\abs{\bold{p}}/ m_{\phi} \ll 1$. In this regime we have that $\epsilon_2 \approx m_{\phi}$ and the logarithmic functions are expanded in the following way 
\begin{align}
    L_2^{\pm}(\abs{\bold{k}}) &\approx \log\left[\frac{  m_{\phi}^2 + 2 m_{\phi} \abs{\bold{p}} + 2 \abs{\bold{p}} \abs{\bold{k}}}{ m_{\phi}^2 + 2 m_{\phi} \abs{\bold{p}} - 2 \abs{\bold{p}} \abs{\bold{k}}}\right] \pm \log\left[\frac{ m_{\phi}^2 - 2 m_{\phi} \abs{\bold{p}} + 2 \abs{\bold{p}} \abs{\bold{k}}}{ m_{\phi}^2 - 2m_{\phi} \abs{\bold{p}} - 2 \abs{\bold{p}}  \abs{\bold{k}}}\right] \nonumber \\ 
    & \approx 
    2 \log\left[\frac{1 + 2 \abs{\bold{k}} \abs{\bold{p}}/ m_{\phi}^2}{1 - 2 \abs{\bold{k}} \abs{\bold{p}}/ m_{\phi}^2}\right] \approx 8 \frac{\abs{\bold{k}}\abs{\bold{p}}}{m_{\phi}^2} + \frac{32}{3} \left(\frac{\abs{\bold{k}}\abs{\bold{p}}}{m_{\phi}^2}\right)^3,
    \label{eq: log expansion 1}
\end{align}
and similarly 
\begin{equation}
     L_1^{\pm}(\abs{\bold{k}})\approx 8 \frac{\abs{\bold{k}}\abs{\bold{p}}}{m_{\phi}^2} + \frac{128}{3} \left(\frac{\abs{\bold{k}}\abs{\bold{p}}}{m_{\phi}^2}\right)^3. 
     \label{eq: log expansion 2}
\end{equation}
By substituting these expressions back into~\eqref{eq: simplified version of Veff} we finally obtain 
\begin{align}
\label{eq:pot2}
    V_{\text{eff}}(\abs{\bold{p}},T_s \ll m_{\phi}) &= -\frac{y^2}{2 \abs{\bold{p}}^2}
 \int_0^\infty \frac{\dd{\abs{\bold{k}}}}{8 \pi^2} \frac{128}{6} \frac{\abs{\bold{k}}^3 \abs{\bold{p}}^3}{m_{\phi}^4}f_f(\abs{\bold{k}})\nonumber \\ &= -\frac{7 \pi^2 y^2 \abs{\bold{p}} T_s^4}{45 m_{\phi}^4} + \mathcal{O}(m_{\phi}^{-5}),
 \end{align} 
where we have again assumed Fermi-Dirac and Bose-Einstein statistics. The results using a Boltzmann distribution are explicitly shown in the main text.
The change of the sign in the potential that is apparent in the comparison of Eq.~\eqref{eq:pot1} and Eq.~\eqref{eq:pot2} can be understood through an analogy between neutrino propagation in a medium with classical optics, see e.g. \cite{Notzold:1987ik}. The potential acts as a medium induced correction to the dispersion relation and can be interpreted as an index of refraction $n_{ref}$ with $V/p= - (n_{ref}-1)$. Neglecting the thermal nature of the medium for the moment, one can also determine $n_{ref}$ from the forward scattering amplitude, which changes sign if there is a resonance in the amplitude. In our case the resonance is from the s-channel exchange of $\phi$ and one expects the potential to change signs from $s<m_\phi^2$ to $s> m_\phi^2$. After averaging over the momenta of the scattering centers this leads to an expected change of sign at $p \times T_s \approx m_\phi^2$ in agreement with the results from the more sophisticated thermal field theory computation. 

\section{Computation of the sterile neutrino rates}
\label{app: Sterile Rates} 

In this appendix we show the details for the calculation of the rates. As noted in the main text, the rate for the sterile neutrinos is comprised by two pieces, each one corresponding either to $\nu_s \nu_s \leftrightarrow \nu_s \nu_s$ or $\nu_s \nu_s \leftrightarrow \phi \phi$. In each case, the general form of the rate is 
\begin{equation}
    \Gamma_{\nu_s \nu_s \leftrightarrow k k}(p_1) = \frac{1}{2 E_1} \int \dd{\Pi}_2 \dd{\Pi}_3 \dd{\Pi}_4 (2 \pi)^4 \delta^4(p_1+p_2-p_3-p_4) \abs{\mathcal{M}}^2_{\nu_s \nu_s \leftrightarrow k k  } f_s(p_2),
    \label{eq: general equation rate}
\end{equation}
where $p_1$ and $p_2$ are the four-momentum of the incoming particles and $p_3$ and $p_4$ the four-momentum associated with the outgoing particles. The Lorentz-invariant phase-space volumes are defined as $\dd{\Pi}_i = \frac{\dd[3]p_i}{(2\pi)^3 2 E_i}$ and the squared matrix elements are
\begin{align}
   \abs{\mathcal{M}}^2_{\nu_s \nu_s \leftrightarrow \phi \phi} &= 4 y^4 \left[\frac{\left(s +2t -2 m_{\phi}^2\right)^2 \left(-m_{\phi}^4 + 2 m_{\phi}^2 t - t(s+t)\right)}{t^2 \left(m_{\phi}^2 -s -t\right)^2} \right] ,
\end{align}
and
\begin{align}
\abs{\mathcal{M}}^2_{\nu_s \nu_s \leftrightarrow \nu_s \nu_s} &= 8 y^4 \left[ \frac{s^2}{(s-m_{\phi}^2 )^2 + m^2_{\phi} \Gamma_{\phi}^2}  - \frac{ (s^2 +st) \left[(s-m^2_{\phi})(-t-s-m^2_{\phi}) \right] }{\left[(s-m_{\phi}^2 )^2 + m^2_{\phi} \Gamma_{\phi}^2\right]\left[(s+t+m_{\phi}^2 )^2 \right]}\right. \nonumber \\ & + \frac{t^2}{(t-m_{\phi}^2 )^2} - \frac{ (t^2+st) \left[(t-m^2_{\phi})(-t-s-m^2_{\phi}) \right] }{\left[(t-m_{\phi}^2 )^2 \right]\left[(t+s+m_{\phi}^2 )^2 \right]}   \nonumber\\ &\left.+\frac{(s+t)^2}{(-s-t-m_{\phi}^2 )^2 } + \frac{ st \left[(s-m^2_{\phi})(t-m^2_{\phi}) \right] }{\left[(s-m_{\phi}^2 )^2 + m^2_{\phi} \Gamma_{\phi}^2\right]\left[(t-m_{\phi}^2 )^2 \right]} \right],
\end{align}
with $s$ and $t$ being Mandelstan variables defined in the usual way and $\Gamma_{\phi}$ being the width of the scalar
\begin{equation}
    \Gamma_{\phi} = \frac{y^2}{8 \pi} m_{\phi}.
\end{equation}
Note that the integral in~\eqref{eq: general equation rate} can be rewritten in terms of the cross section 
\begin{equation}
   \Gamma_{\nu_s \nu_s \leftrightarrow k k}(p_1) = \frac{1}{2 E_1} \int \dd{\Pi}_2 f(p_2)~ 4 (p_1 \cdot p_2) ~  \sigma_{\nu_s \nu_s \leftrightarrow k k} 
   \label{eq: simplified rate}
\end{equation} 
where 
\begin{align}
    \sigma_{\nu_s \nu_s \leftrightarrow k k} &= \int \dd{\Pi}_3 \dd{\Pi}_4 (2 \pi)^4 \delta^4(p_1+p_2-p_3-p_4) \frac{\abs{\mathcal{M}}^2_{\nu_s \nu_s \leftrightarrow k k}}{4~ (p_1 \cdot p_2)},
\end{align}
and we have assumed ultra-relativistic sterile neutrinos, \emph{i.e.} $E\approx \abs{\bold{p}}$. The cross sections for the processes discussed in the main text read 
\begin{align}
    \sigma_{\nu_s \nu_s \leftrightarrow \nu_s \nu_s} =& \frac{y^4}{2 \pi s^2((m_{\phi}^2-s)^2+m_{\phi}^2 \Gamma_{\phi}^2)} \left[ \frac{s(5 m_{\phi}^6 -9m_{\phi}^4 s +6s^3)}{m_{\phi}^2 +s} +  \frac{2(5 m_{\phi}^8 - 9 m_{\phi}^6 s + 4s^3)\log\left( \frac{m_{\phi}^2}{m_{\phi}^2+s}\right)}{2m_{\phi}^2+s} \right],\\ \nonumber 
    \sigma_{\nu_s \nu_s \leftrightarrow \phi \phi} =& \frac{y^4}{4 \pi s^2 }\left[\frac{6m_{\phi}^4 - 4m_{\phi}^2 s +s^2}{2m_{\phi}^2 -s} 2 \log\left(\frac{(s(s-4m_{\phi}^2))^{1/2} +s - 2m_{\phi}^2}{(s(s-4m_{\phi}^2))^{1/2} -s + 2m_{\phi}^2}\right) - 6(s(s-4m_{\phi}^2))^{1/2}\right].
\end{align}
Our results for the cross sections match those reported for the same model in a different context in \cite{Esteban:2021tub,Doring:2023vmk}.
The remaining integral over $p_2$ in~\eqref{eq: simplified rate} must be solved numerically. Nonetheless, for the process $\nu_s \nu_s \leftrightarrow \nu_s \nu_s$ one can find an analytical expression for the rate in the `heavy' ($m^2_{\phi} \gg p T$) and the `massless' limit ($m^2_{\phi}\ll pT$), if one assumes a Boltzmann distribution for $f(p_2)$
\begin{equation}
   \Gamma_{\nu_s \nu_s \leftrightarrow \nu_s \nu_s}(p) =  \begin{cases}
        \frac{3 y^4 T_s^2}{2 \pi^3 p } e^{\frac{\mu}{T_s}} & p T_s \gg m^2_{\phi} \\[10pt]
        \frac{20 y^4 p  T_s^4}{3 \pi^3 m_{\phi}^4} e^{\frac{\mu}{T_s}} & p T_s \ll m^2_{\phi}.
    \end{cases}
\end{equation}
Moreover, for intermediate temperatures where the rate is dominated by the s-channel resonance, 
one can use the narrow width approximation and the integral is exact
\begin{equation}
\Gamma_{\nu_s \nu_s \leftrightarrow \nu_s \nu_s}^{\text{resonant}}(p) = \frac{y^2 T_s m_{\phi}^2}{2 \pi p^2}e^{-\frac{m_{\phi}^2}{4 p T_s}+\frac{\mu}{T_s}}. 
\end{equation}
The $\nu_s \nu_s \leftrightarrow \nu_s \nu_s$ rate is, therefore, well approximated by a piece-wise function
\begin{equation}
   \Gamma_{\nu_s \nu_s \leftrightarrow \nu_s \nu_s}(p) =  \begin{cases}
        \frac{3 y^4 T_s^2}{2 \pi^3 p } e^{\frac{\mu}{T_s}} +\frac{y^2 T_s m_{\phi}^2}{2 \pi p^2}e^{-\frac{m_{\phi}^2}{4 p T_s}+\frac{\mu}{T_s}} & p T_s > \frac{3 m_{\phi}^2}{2\sqrt{10}} \\[10pt]
        \frac{20 y^4 p  T_s^4}{3 \pi^3 m_{\phi}^4} e^{\frac{\mu}{T_s}} +\frac{y^2 T_s m_{\phi}^2}{2 \pi p^2}e^{-\frac{m_{\phi}^2}{4 p T_s}+\frac{\mu}{T_s}} &  p T_s \leq \frac{3 m_{\phi}^2}{2\sqrt{10}}.
    \end{cases}
\end{equation}
In \cite{Doring:2023vmk} a comparison of the full numerical result for the rate with a similar analytic approximation is shown. The is already very good for the rather large Yukawa $y=0.2$ shown there and corrections are expected to be even smaller in the parameter space considered in our work. As the corrections are related to the narrow width approximation we expect them to be of order $\Gamma_{\phi}/m_\phi$.

\begin{figure}[tbp]
\centering
\includegraphics[width=.7\textwidth]{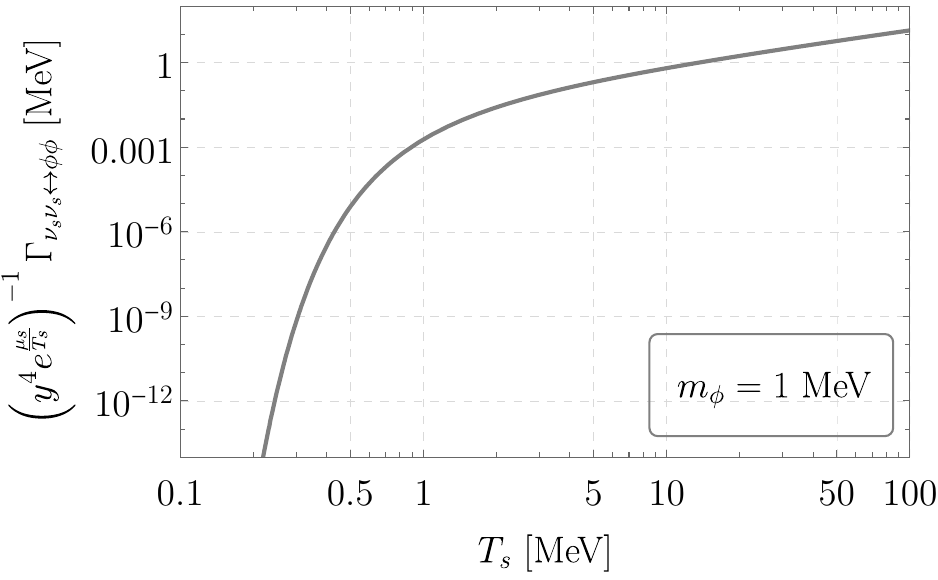}
\caption{\label{fig: rate 2 to 4} Dependence of $\left(y^4 e^{\frac{\mu_s}{T_s}} \right)^{-1} \Gamma_{\nu_s \nu_s \leftrightarrow \phi \phi} $ with respect to the steriles' temperature $T_s$ for a fixed mediator mass $m_{\phi}= 1~\text{MeV}$ and for neutrinos with a typical momentum $p\sim T_s$.} 
\end{figure}

On the other hand, the rate for the $\nu_s \nu_s \leftrightarrow \phi \phi$ process has no simple analytical expression so it must be evaluated numerically. Its dependence on the temperature is shown in Fig.~\ref{fig: rate 2 to 4} for an exemplary mediator mass of $m_{\phi} = 1 ~\text{MeV}$. As expected, for high temperatures the rate shows a scaling $\Gamma_{\nu_s \nu_s \leftrightarrow \phi \phi} \sim T_s$, while it is exponentially suppressed once $T_s \lesssim m_{\phi}.$  

\bibliographystyle{JHEP.bst}
\bibliography{sterile_nu_DM.bib}
\end{document}